\documentclass{siamltex}
\usepackage{amsmath}
\usepackage{amssymb}
\usepackage{bm}
\usepackage{algorithmicx}
\usepackage{algorithm} 
\usepackage[noend]{algpseudocode} 
\usepackage{bbm}
\usepackage{graphicx}
\usepackage{url}
\usepackage{color}
\usepackage[section]{placeins}
\usepackage{subcaption}

\title{Rigid graph compression: Motif-based rigidity analysis for disordered fiber networks\thanks{This work was supported by the U.S. Army Research Office (\#W911NF-13-1-0013 and \#W911NF-16-1-0356). Additional support was provided by a James S. McDonnell Foundation 21st Century Science Initiative - Complex Systems Scholar Award (\#220020315) and by the Eunice Kennedy Shriver National Institute of Child Health \& Human Development of the National Institutes of Health (R01HD075712).
The content is solely the responsibility of the authors and does not necessarily represent the official views of any of the agencies that supported this work.
}
}

\author{
Samuel Heroy\thanks{Carolina Center for Interdisciplinary Applied Mathematics, Department of Mathematics, University of North Carolina, Chapel Hill, NC 27599, USA (sam.heroy@gmail.com, dane.r.taylor@gmail.com, forest@unc.edu, mucha@unc.edu)} 
\and Dane Taylor\footnotemark[2]
\thanks{Department of Mathematics, University at Buffalo, State University of New York, Buffalo, NY 14260, USA}
\and F. Bill Shi\thanks{The Odum Institute for Research in Social Science, University of North Carolina, Chapel Hill, NC 27599, USA (billanson10@gmail.com)}
\and M.~Gregory Forest\footnotemark[2]
\thanks{Department of Applied Physical Sciences, University of North Carolina, Chapel Hill, NC 27599, USA }
\and Peter J. Mucha\footnotemark[2] \footnotemark[5]
}

\begin{document}

\maketitle 

\begin{abstract}
Using particle-scale models to accurately describe property enhancements and phase transitions in macroscopic behavior is a major engineering challenge in composite materials science. 
To address some of these challenges, we use the graph theoretic property of rigidity to model mechanical reinforcement in composites with stiff rod-like particles. 
We develop an efficient algorithmic approach called \emph{rigid graph compression} (RGC) to describe the transition from floppy to rigid in disordered fiber networks (`rod-hinge systems'), which form the reinforcing phase in many composite systems.
%
%
To establish RGC on a firm theoretical foundation, we adapt rigidity matroid theory to identify primitive topological network motifs that serve as rules for composing interacting rigid particles into larger rigid components.
This approach is computationally efficient and stable, because RGC requires only topological information about rod interactions (encoded by a sparse unweighted network) rather than geometrical details such as rod locations or pairwise distances (as required in rigidity matroid theory).
We conduct numerical experiments on simulated two-dimensional rod-hinge systems to demonstrate that RGC closely approximates the rigidity percolation threshold for such systems, through comparison with the pebble game algorithm (which is exact in two dimensions).
Importantly, whereas the pebble game is derived from Laman's condition and is only valid in two dimensions, the RGC approach naturally extends to higher dimensions.
\end{abstract}

\begin{keywords}
{fiber networks, composite materials, rigidity, graph compression, network motifs, rigidity matroid theory, pebble game}
\end{keywords}

\begin{AMS}
60K35, 68R10, 82B43, 90C27, 91D25, 94C15, 	05C62, 05C85
\end{AMS}

\pagestyle{myheadings}
\thispagestyle{plain}
\markboth{S. HEROY {\it et al.}}{RIGID GRAPH COMPRESSION}

\section{Introduction}
High aspect ratio particles (e.g., thin rods) of nanoscopic or microscopic scales are routinely incorporated into polymeric host materials to enhance attributes such as electrical and thermal conductivity, charge storage, and mechanical resilience. These composites often exhibit a nonlinear response with respect to the density (measured by volume fraction, $\phi$) of rods or other filaments: the property gain scales linearly at small volume fractions, then increases dramatically as $\phi$ passes through a critical threshold $\phi_c$. When considering the conductivity of a poorly conducting polymer enhanced with highly conductive rods, this sharp transition is associated with `contact percolation,' wherein interacting rods form a giant and spatially extended network component (see, e.g., \cite{shi}). An analogous sharp rise in mechanical stability at volume fractions greater than or equal to $\phi_c$ has been observed in numerous composite systems \cite{MPCellulose, MPCellulose2,lithium,graphene,buckypaper}. However, characterization of the physical mechanism underlying `mechanical percolation' remains an open problem---no emergent phenomenon (akin to contact percolation) in rod or rod-polymer interactions has been shown to trigger this macroscopic behavior.

The nature of mechanical percolation varies considerably with the complexity of the composite system. As an example, cellulose fibers (or whiskers) obtained from tunicin have high tensile modulus ($\sim$120-150 GPa) and high aspect ratio (10-20 nm width and 100 nm to several $\mu $m length) \cite{MPCellulose, MPCellulose2}. In cellulose fiber-reinforced composites, generic percolation models with fitting parameters tuned to data accurately describe the relationship of different moduli to the volume fraction $\phi$ of cellulose fibers \cite{MPCellulose,chanzy,ouali}. The reactivity of readily available hydroxyl groups along the whiskers makes hydrogen bonding interactions especially favorable, such that contacts transmit bending modes in addition to compression modes, thereby fixing the angles between contacting particles \cite{MPCellulose, cellulose}. Thus, the presence of a spatially extended network component of contacting particles is posited to drastically increase the stiffness of these composite materials, consistent with experimental results as well as with physically-based two-dimensional simulations \cite{MPCellulose, MPCellulose2,stiffpolymer}. In other systems, however, attractive forces are relatively soft and only transmit compressive and tensile forces, so that contact percolation alone is not enough to mechanically stabilize the network and additional constraints are needed. In such systems, experiments show that mechanical percolation occurs at higher volume fractions than electrical percolation, both when the reinforcement particles are of high aspect ratio \cite{compressed_graphite,lithium,rheological}, and otherwise \cite{niklaus,graphene}.

The mechanical properties of any composite material depend on the specific properties of each phase, the volume fraction as well as morphology of the reinforcing components, and the interfacial properties \cite{cellulose}. Homogenization models, such as the Halpin-Tsai equations, have been successfully adapted to a variety of systems with different morphologies and interfacial properties to predict modulus as a function of volume (or weight) fraction of the reinforcing phase \cite{compressed_graphite,nanoreview,H-T}. Micromechanical modeling efforts have provided more sophisticated and accurate models, which take into account the interplay between random microstructure and interphase \cite{baxter2,BAXTER,fralick,Riu-Brinson}. While many different classes of mechanical models are generally useful and have more immediate predictive capability than our study here, none of these provide an explicit mechanism for the emergent nonlinear gain in mechanical properties generated by favorable interactions between particles within the reinforcing phase. Regarding these interactions, a high aspect ratio is presumably advantageous as it allows a higher number of contacts at a fixed volume fraction---if these contacts are sufficiently attractive, then the emergent network of contacts will contribute significantly to the composite mechanical properties.

\begin{figure}[t]
\centering
\includegraphics[width=.8\linewidth]{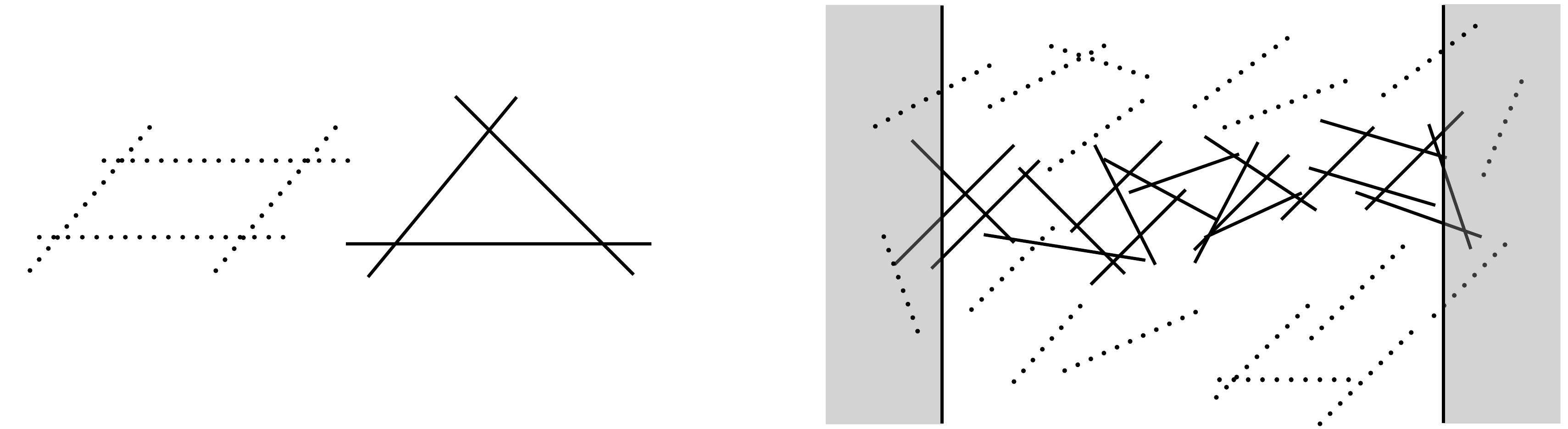}
\caption{
{\bf Rigidity in two-dimensional rod-hinge systems.} Left: Supposing that rods interact as hinges at intersections, four rods connected pairwise in two dimensions are `floppy' (dotted) in that they may deform through different interior angles, whereas three rods connected pairwise are `rigid' (solid). Right: A large component of mutually rigid rods may add mechanical stability to a host composite. We use rigidity characterization algorithms to detect the presence of such a component connecting two boundaries (vertical rods) bounding a large computational domain. In order to keep the rod distribution uniform throughout the (white) domain, we allow rods to be placed in the `buffer' regions (grey rectangles) on the exterior sides of the boundaries.}
\label{fig:rods1}
\end{figure}

It is our hypothesis that mechanical percolation occurs when the reinforcing phase has sufficient volume fraction so as to coalesce into a giant \emph{rigid scaffold}, wherein the individual reinforcing particles are not only in contact, but furthermore the constraints that result from these connections are sufficient to eliminate any nontrivial degrees of freedom (`floppy modes') within the scaffold (see Figure \ref{fig:rods1}). Our graph-theoretic approach to modeling composite mechanical properties is motivated in part by successes using a similar approach for modeling conductance in rod-based composites. Using a simple model in which randomly dispersed conductive rods interact solely through contacts with one another, Shi \emph{et al.} both detect and determine the consequences of contact percolation under a variety of rod dispersion anisotropies (evoking realistic processing conditions) \cite{scaling,shi,xiaoyu}, robustly capture various multi-scale properties of the composite, including the exponential tails of current distributions above the contact percolation threshold.  We use a similar network representation to characterize mechanical properties, assuming that stiff rods interact solely through pairwise attractive contacts in an otherwise soft medium.

As discussed in Sec.~\ref{sec:back}, a number of studies have been devoted to the characterization of \emph{rigidity} and \emph{rigidity percolation}, but none of the associated methods have proven adequate for large three-dimensional rod systems. Unlike previous studies of rigidity percolation, our work takes the perspective that large rigid scaffolds are `built up' from simple topological patterns that apply at any scale. With this perspective, we identify and prove primitive \emph{rigid motifs}---specific contact rules that determine when a small set of interacting rigid components are together rigid---so that we may agglomerate these motifs into larger rigid components. In two dimensions, these motifs do not require specification as to whether the separate components be individual particles or aggregates themselves; and only minimal specification as to whether these particles be rods, ellipsoids, curvilinear fibers, whiskers, etc. In three dimensions, the problem is only slightly more complicated in that cylindrical particles that have one less degree of freedom (5) than more complex (asymmetrical) rigid bodies (6). The algorithm we develop using this `topological building blocks' perspective---\emph{rigid graph compression} (RGC)---is therefore applicable to any number of spatial dimensions upon selection of appropriate motifs, and is also applicable to systems of varying particle shape (nanorods, ellipsoids, curvilinear fibers, etc.). The present study develops and demonstrates the feasibility and accuracy of rigid graph compression in two-dimensional \emph{rod-hinge systems}, leaving the consideration of and application to three-dimensional systems for subsequent study.

We use the term `rod-hinge systems' to denote physical networks of randomly distributed, high aspect ratio, and completely inflexible cylinders (which may be used to model stiff fibers, rods, whiskers, nanotubes, etc.). In two dimensions, these rods may be infinitesimally thin line segments which physically intersect (`Mikado models' \cite{head,head2,stiffpolymer}), while in three dimensions the cylinders must have finite radii to intersect. We model contacts between cylinders as hinges, such that intersecting rods may pivot about them but assume that friction or other forces keep the rods in contact at these points. When an ensemble of rods has enough contacts such that these constraints keep all rods fixed relative to each other, they may only move as a single \emph{rigid component}, and we refer to the rods in this body as being mutually rigid. In Sec.~\ref{sec:back}, we discuss existing methods to characterize rigidity, as well as their strengths and drawbacks for modeling rod-hinge systems. In Sec.~\ref{sec:main}, we prove simple topological rules regarding how \emph{motifs} of rigid components connect to become a larger rigid component, which we then incorporate into our RGC algorithm. In Sec.~\ref{sec:results}, we use RGC to estimate the rigidity percolation threshold and correlation length exponent in random two-dimensional rod-hinge systems, comparing with previously available methods.

\section{Previous Rigidity Characterization Methodologies}\label{sec:back}
We begin with a review of existing work on rigidity characterization in random systems. Such work is typically framed in the context of central force (CF) networks, simple graphs $G(V,E)$ embedded in $D$-dimensional Euclidean space wherein the set of edges (`bonds'), $E$, denotes fixed distances between members of the vertex set, $V$ (`nodes' or `atoms') \cite{laman,hendrix}. Importantly, the rod-hinge systems of interest here include additional constraints beyond their representation as CF networks; but for completeness we nevertheless start with this review of methods used for CF networks. We describe Maxwell's Isostatic Condition, a global condition for rigidity analysis (Sec.~\ref{sec:maxwell}); graph theoretic methods based on {Laman's Condition} for both the deconstruction and construction of rigid networks (Sec.~\ref{subsec:laman}); and finally a dynamical systems approach to rigidity analysis in {rigidity matroid theory} (Sec.~\ref{sec:rmt}). In Sec.~\ref{sec:previous_discussion}, we discuss the limitations of these methods for rigidity characterization in large rod-hinge systems. 

\subsection{Maxwell's Isostatic Condition}\label{sec:maxwell}

Purely topological characterization of the rigidity of graphs can be traced back to Maxwell's isostatic condition, which claims that a graph is rigid if the number of constraints, $|E|$, is equal to the inherent number of degrees of freedom, $D|V|$ \cite{maxwell}. 
Latva-Kokko and Timonen applied this condition to rigidity percolation in 2D `random networks of stiff fibers' (rod-hinge systems), with density of unit-length rods $q=r/A$, where $r$ is the number of rods, and $A=L^2$ is the dimensionless system area \cite{pebgame2drods2,pebgame2drods,statgeoclassic}. In this system, the number of contacts $N_c$ scales linearly with the total number and density of rods \cite{paper}:  
\begin{align}
N_c\approx rq/\pi
\label{crosspoints}
\end{align}
A contact between two rods constrains two degrees of freedom. Assuming without justification (and incorrectly) that all such constraints are independent of one another, one infers that the network becomes rigid at the rigidity percolation threshold $q_\mathrm{min}$ satisfying
\begin{align}
3r=2N_c=\frac{2rq_\mathrm{min}}{\pi}\Rightarrow q_\mathrm{min}=\frac{3}{2}\pi \doteq 4.71\,.
\label{lvpredict}
\end{align}

As we will show in Sec.~\ref{sec:results}, the density predicted for rigidity percolation based on Maxwell's isostatic condition in this system is far too low---indeed, it is even lower than the contact percolation threshold for two-dimensional isotropically random rod systems \cite{statgeoclassic} ($q_c\doteq 5.71$). Importantly, the number of contacts per rod obeys a Poisson distribution, and thus many constraints redundantly bind the same free motions within the system (`floppy modes'), while others are left unconstrained \cite{Thorpe2}.

\subsection{Laman, the Pebble Game, and Henneberg Constructions}\label{subsec:laman}
The inadequacy of Maxwell's condition in random systems necessitates a description accounting for dependence between constraints. \emph{Laman's condition} can be used to determine when the constraints in a 2D central force network are independent \cite{laman}. 
\begin{theorem}[Laman's Condition]
 The edges of a graph $G(V,E)$ are independent in two dimensions if and only if no subgraph containing $V'$ vertices has more than $2V'-3$ edges. If a graph has exactly $2V'-3$ edges within each subgraph on $V'$ vertices, then it is called a Laman graph.
\label{Laman}
\end{theorem}

Rigid graph characterization based on direct application of Laman's Condition would require iterating tests upon every subgraph, which would be computationally hopeless for all but the smallest systems. However, an equivalent formulation of this condition is the following: the edges of $G(V,E)$ are independent in two dimensions if and only if, for each edge $e_{ij}\in E$, the graph formed by adding three new edges between $i$ and $j$ has no subgraph on V' nodes with more than 2V' edges. Jacobs and Thorpe developed this corollary into a ``pebble game" test for independence of edges with computational complexity that scales in the worst case as ${\cal O}(|V|^{1.2})$ \cite{rmtheory,pebble}. In 2D CF networks, each node is assigned a pair of `pebbles' to represent the two degrees of freedom of a point in a plane. Each bond between these nodes pins down one of their allotted pebbles, so long as there are sufficient pebbles for each bond---otherwise, some edges are redundant, or `stressed.'  Once the pebble game is implemented, the locations of free pebbles and enumeration of independent edges allow for decomposition of the network into sets of mutually rigid nodes, alongside an accurate count of the system's net degrees of freedom. The pebble game has been applied to a variety of systems, including CF networks with random topologies (i.e., Erd\H{o}s R\'{e}nyi graphs), wherein Thorpe \emph{et al.} show that networks undergo a rigidity transition as the mean coordination number (average degree) approaches $\approx 4$  \cite{glasses}.

The pebble game is deconstructive in that it is used for partitioning a graph into rigid and floppy components. An alternative goal is the construction of rigid graphs, which can be accomplished for 2D CF networks using \emph{Henneberg constructions}, inductive rules for the construction of Laman graphs \cite{henneberg}. Constructions begin with an edge connecting two vertices (a trivially rigid graph). Then, the following steps are repeated iteratively: a new vertex is added, adjoined either (a) to two vertices via two new edges; or (b) to two previously adjacent nodes, while the old edge between these latter nodes is severed and a third edge is placed between the new node and another previously existing node. As noted above, every Henneberg construction is a Laman graph; but perhaps more surprisingly, every Laman graph can be realized by Henneberg constructions \cite{tay}.

\subsection{Rigidity Matroid Theory}\label{sec:rmt}
Unlike the approaches described above, rigidity matroid theory uses a graph's embedding in Euclidean space, or `framework', $\rho(G)$, to characterize its rigidity through the language of linear algebra \cite{singer,graver, hendrix}.
Consider the set of node positions to be a dynamical system such that $\bm p_i(t)$ is the $D$-dimensional position of node $i$ at time $t.$ The condition that each edge $e_{ij}\in E$ maintains a fixed distance $d_{ij}$ between nodes $i$ and $j$ requires $\sum_{D'=1}^D|p_i^{D'}(t)-p_j^{D'}(t)|^2=d_{ij}^2\;\;\forall e_{ij}\in E$.
Since this quadratic system is not computationally convenient, it is common to linearize by differentiating with respect to time, obtaining
\begin{equation}
(\bm p_i(t)-\bm p_j(t))\cdot(\bm u_i(t)-\bm u_j(t))=0
\qquad
\forall e_{ij}\in E\,,
\label{rmt}
\end{equation} 
where $\bm u_i(t)={d\bm p_i(t)}/{dt}$ is the instantaneous velocity of node $i$. The totality of these constraints informs an $|E|\times D|V|$ matrix, $\bm X$---the rigidity matrix of $\rho(G)$---satisfying $\bm X \bm u= \bm 0$, where $\bm u$ is the $D|V|$-vector of velocities. 
A vector $\bm u$ satisfying $\bm X \bm u= \bm 0$ is an \emph{infinitesimal motion} of $\rho(G)$, and the right nullspace of $\bm X$ includes the full set of such motions. 
If $G$ is embedded in Euclidean $\mathbb{R}^D$ and the right nullspace of $\bm X$ spans only the $D(D+1)/2$ rigid-body motions of translation and rotation, the framework $\rho(G)$ is said to be \emph{infinitesimally rigid}. Otherwise, $\rho(G)$ is \emph{infinitesimally flexible}.

Importantly, it has been shown generically that if a framework $\rho(G)$ is infinitesimally rigid, then all other realizations of $G$ are infinitesimally rigid \cite{Gluck}. 
Therefore, one can (generically) infer rigidity from the topology of a graph itself, rather than from any particular embedding in space.
We note that this argument breaks down when at least one nontrivial minor of $\bm X$ has a zero determinant---however, these cases occur with probability zero in random systems. Practically, determining the rigidity of $\rho(G)$ thus reduces to computing the rank of $\bm X$, and using the rank nullity theorem to then determine the dimension of the matrix's null space, which corresponds injectively to the underlying graph's degrees of freedom.


\subsection{Relationship to Rod-Hinge Systems}\label{sec:previous_discussion}

These prior methodologies have been central to the study of graph rigidity; however, none are entirely suited to the rigidity analysis of rod-hinge systems.
First, the techniques described in Secs.~\ref{subsec:laman}--\ref{sec:rmt} have been developed for CF networks; however, in rod-hinge systems, the position of each contact between two rods is fixed relative to the entirety of both of these rods, which may include any number of contacts with other rods. Refs.~\cite{pebgame2drods2,pebgame2drods} introduce augmented constraints between second nearest neighbors to extend the pebble game to 2D rod-hinge systems, and use this extended method to characterize the critical rod density of rigidity percolation as well as the scaling of rigid component size near the threshold. 
%
However, because it depends on Laman's condition for 2D graphs, the pebble game cannot extend exactly to three dimensions in its current form, and it is unclear whether any such breadth first search algorithm could even possibly account for the complicated variety of three dimensional floppy modes. Instead, the pebble game has been only approximately extended to 3D glass-like networks \cite{threedpebble}; it is unclear whether or not this approximate extension can be modified for 3D rod-hinge systems, and if it indeed can, what accuracy it might obtain.

One can instead augment rigidity matroid theory for rod-hinge systems (see App.~\ref{app:motif_proofC}). However, while rigidity matroid theory is (unlike the pebble game) valid in any number of dimensions, rigidity matrices do not offer any local information about which sets of rods are rigid relative to others. That is, while one could in principle use rigidity matrices to tally the macroscopic degrees of freedom in various large systems, they would not immediately characterize rigidity percolation (barring an extremely exhaustive brute force search of submatrices) \cite{hendrix}. Furthermore, because rigidity matrices rely on the full set of rod intersection points, many of these points will be spatially close at sufficiently high rod densities, subjecting the analysis of the corresponding matrix to numerical error \cite{hendrix,singer}. 

Finally, we consider another previously used technique for rigidity analysis, in which rods are considered to be stiff springs that connect at rod intersection points \cite{stiffpolymer,head,head2}. These points are first subjected to a perturbation (corresponding to physical deformation), and then the spring system is relaxed using nonlinear optimization. If the initial pairwise distance between two nodes is maintained after relaxation, then the points (and the rods containing them) are deemed rigid with respect to one another; otherwise, they are not. This method does allow for characterization of rigidity percolation---finding similar results to those of the pebble game \cite{stiffpolymer})---but being a search for the ``lowest point of a complicated high-dimensional valley with extremely steep slopes but hardly varying base altitude'' \cite{stiffpolymer}, it was observed to be highly unstable in large 2D systems and has not been attempted in 3D systems.
	
To help guide our development of scalable algorithms for rigidity analyses in both two dimensions and beyond, we highlight that there is a hierarchy of information required across these previous methods. Maxwell's isostatic condition requires only the global density of rods ($q$); the pebble game and Henneberg constructions require topological information, specifying which \emph{contact points} are adjacent (i.e., an edge list or adjacency matrix); rigidity matroid theory and the spring relaxation method additionally require the spatial locations of rod intersection points. In the next section, we derive an alternative method that stems from rigidity matroid theory, and can therefore be generalized to higher dimensions. However, in application this method is an efficient topological algorithm that only requires a list of rod contacts instead of the full knowledge of the locations of contacts.

\section{Motif-Based Rigidity Decomposition}\label{sec:main}

We now introduce Rigid Graph Compression (RGC): a multi-scale, motif-based approach to rigidity analysis of rod-hinge systems that can in principle be extended to any number of dimensions, as well as to systems of different particle shapes. To simplify our presentation and focus on systems where RGC can be compared to exact methods for rigidity analysis, we here restrict our attention to 2D rod-hinge systems, applying our methodology to 3D in future work. Our methodology includes three main parts: in Sec.~\ref{sec:geometry}, we apply rigidity matroid theory to characterize rigid components using an appropriate number of linear constraints. 
In Sec.~\ref{sec:motifs}, we use this foundation to identify primitive rigid motifs that will serve as the building blocks of large rigid networks. In Sec.~\ref{sec:rgc}, we incorporate these primitive motifs into our scalable RGC algorithm.

\subsection{Rigidity Matroid Theory for Interacting Rigid Components}\label{sec:geometry}~
We use rigidity matroid theory (Sec.~\ref{sec:rmt}) to study the motions of small numbers of interacting rigid components. The motions of any rigid component in $D$ dimensions can be fully determined from $D+1$ points contained in the component, the translations and rotations of which together generate the Special Euclidean group \emph{SE(D)} \cite{geometries}. (In principle, fewer coordinates are needed if employing angular constraints, but for simplicity we work with $D+1$ points.) Hence, for some rigid body $R\subset\mathbb{R}^D$, defined as the union of volumes enclosed within some integer number $r>0$ of $D$-dimensional rods, we affix a \emph{coordinate labeling} of $R$ composed of at least $D+1$ points $\{\bm p_i\}$ which fully capture the rigid motions of $R$. Importantly, no more than two points in a coordinate labeling may be collinear, or else the coordinate labeling will only capture a subset of the rigid motions of the corresponding body. Due to the rigidity of $R$, the pairwise distances between the points are fixed, providing ${D+1}\choose{2}$ constraint rows in the corresponding rigidity matrix $\bm X$, each having the form:
\begin{equation} 
\Delta \bm p_{i,j} \cdot \bm u_{i} -\Delta \bm p_{i,j} \cdot \bm u_{j}=0, 
\label{eq:rigidity_row}
\end{equation} 
where $\bm u_i$ and $\bm u_j$ are the instantaneous velocities corresponding respectively to points $\bm p_i$ and $\bm p_j$ affixed in $R$ and $\Delta \bm p_{i,j}=\bm p_i-\bm p_j$. Note that regardless of the dimension of the rod network, if $R$ includes only a single rod, then it has a spatial dimension $1$, and so only two points are needed to specify its rigid motions.

\begin{figure}[t!]
\begin{center}
\includegraphics[width=.9\linewidth]{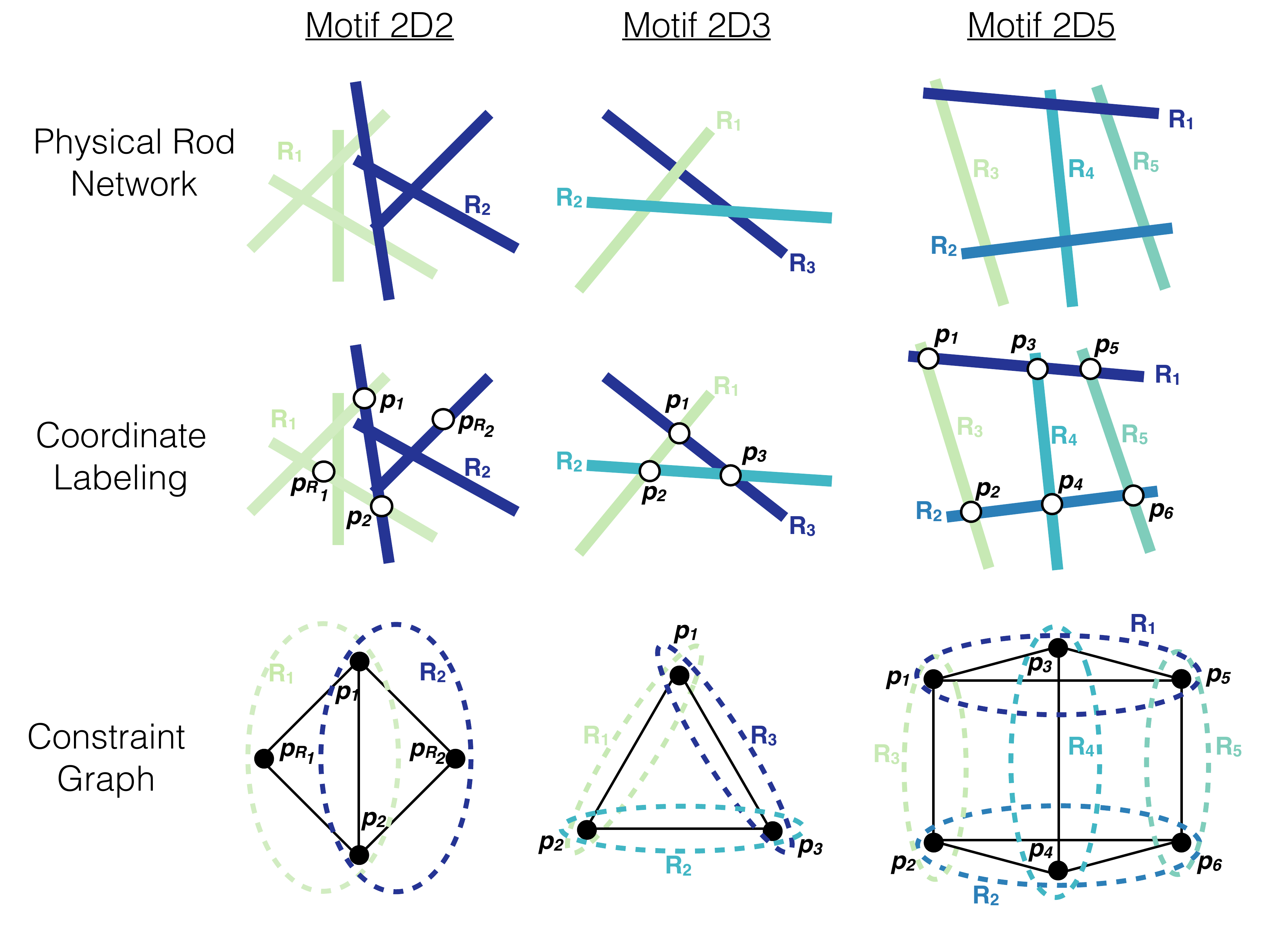}
\caption{{ {\bf Derivation of three primitive rigid motifs for 2D rod-hinge systems.}}
Top Row: Rigid components, which may be individual rods or sets of connected rods, distinguished here by color, intersect with three specific topologies as described in Sec.~\ref{sec:motifs} to form larger-scale rigid components:
(left column) two rigid components $R_1$ and $R_2$ interacting at a pair of points;
(middle column) three rigid components $R_1$, $R_2$ and $R_3$ interacting pairwise; 
and (right column) five rigid components, $R_1,\dots,R_5$ interacting in an identified pattern.
For simplicity, we depict the rigid bodies in the middle and right columns as rods, but the proofs are general to include composite rigid components.
Middle Row: Coordinate labelings are affixed to each rigid component: three noncollinear points are required to describe the motions of a 2D rigid component consisting of multiple rods, whereas individual rods are 1D and require only two points (although more may be used).
For each motif, we identify a set of minimal coordinate labelings that include intersection points whenever possible (see text for clarification).
Bottom Row: The coordinate labelings give rise to constraint graphs in which edges (black lines) indicate distances between adjacent points that are fixed. The dashed ellipses group the rigid components to which these points belong. By Theorems~\ref{axiom1_simple}, \ref{thm:2D3}, and \ref{thm:2D5}, these constraint graphs and the motifs that generated them are rigid in two dimensions.}
\label{fig:axioms}
\end{center}
\end{figure}

When two or more rigid components $\{R_i\}$ interact in contact, we denote the composite system by $R_1* R_2*\dots$ and the corresponding composite rigidity matrix by $\bm X_1 * \bm X_2*\dots$. For such systems, we construct sets of \emph{minimal coordinate labelings}, defined as a union of coordinate labelings for $\{R_i\}$ such that coordinate labelings include points of interaction between rigid components wherever possible. For a given set of coordinate labelings of $\{R_i\}$, we construct a \emph{constraint graph} encoding the topology of interactions (i.e., physical constraints) between the rigid components. This graph is constructed by creating for each rigid component $R_i$ with $S_i$ 
coordinate labels a $(S_i)$-clique---that is, an all-to-all connected subgraph. The constraint graph is defined as the union of these cliques.

\subsection{Primitive Rigid Motifs in 2 Dimensions}\label{sec:motifs}
We now use the above description of rigid components to develop rules for aggregating rigid components into larger composite rigid components. These rules are expressed in three \emph{primitive rigid motifs} (Fig.~\ref{fig:axioms}), which represent our topological `building blocks' of rigidity in 2D rod-hinge systems. We use the term `primitive' because these motifs may not be decomposed into simpler motifs, yet many larger, more complicated patterns of interaction can be constructed from these motifs, analogous to the formation of Laman graphs from Henneberg constructions. For the present study, our rigid components are rods or sets of connected rods, but there is very little about the formalism requiring that the individual particles be rods, and the results of this subsection may with only slight modification be extended to ellipsoids, curvilinear filaments, and other 2D shapes, so long as the interactions between the particles are hinge-like. 

To simplify our discussion, we adopt the naming schema ``\emph{Motif} $xDy$" with $x$ indicating the spatial dimension and $y$ indicating the number of aggregating rigid components in the motif. We identify here three primitive rigid motifs of 2, 3, and 5 aggregating components, respectively, proving here that \emph{Motif 2D2} is rigid and outlining the similarly-constructed proofs for \emph{Motifs 2D3} and \emph{2D5} in the appendices.

\begin{theorem}[Motif 2D2]\label{axiom1_simple} 
The composition of two rigid components $R_1$ and $R_2$  that intersect at two or more points $\bm p_1, \bm p_2,...$ in two dimensions is rigid. 
\begin{proof}
As a first case, we assume both rigid components are inherently 2D (that is, $r_1,r_2>1$) and represent each aggregating rigid component using a coordinate labeling with three noncollinear points. Importantly, we require for both coordinate labelings that two of these points, $\bm p_1$ and $\bm p_2$, be the intersection points between the rigid components so that the composite labeling is minimal (if there are more than two intersection points, we pick two arbitrarily). We denote the remaining two coordinate labeling points for $R_1$ and $R_2$ as $\bm p_{R_1}$ and $\bm p_{R_2}$, respectively, and choose them arbitrarily, subject to being noncollinear with $\bm p_1$ and $\bm p_2$, from the sets 
$R_1\backslash R_2$ and $R_2\backslash R_1$ (that is, in the restriction of the space $R_1$ to points not in $R_2$, and vice versa). 
Thus, the coordinate labelings for $R_1$ and $R_2$ are the sets $\{ \bm p_1, \bm p_2, \bm p_{R_1}\}$ and $\{\bm p_1, \bm p_2, \bm p_{R_2}\}$, respectively. (See, for example, the coordinate labeling in Fig.~\ref{fig:axioms}.)

We determine the rigidity of the composite system $R_1 * R_2$ through the rigidity matrix $\bm X_1\ast \bm X_2$, obtained by combining the rigidity matrices of the individual rigid components:
\begin{align}
\bm X_1 =\left[
\begin{array}{ccc}
\Delta \bm p_{1,2} &  -\Delta \bm p_{1,2} & \bm 0 \\
\Delta \bm p_{1,R_1}& \bm 0 & -\Delta \bm p_{1,R_1} \\
\bm 0 &\Delta \bm p_{2,R_1} & -\Delta \bm p_{2,R_1} 
\end{array}\right],\\
\bm X_2 =\left[
\begin{array}{ccc}
\Delta \bm p_{1,2} &  -\Delta \bm p_{1,2} & \bm 0 \\
\Delta \bm p_{1,R_2} & \bm 0  &-\Delta \bm p_{1,R_2} \\
\bm 0 &\Delta \bm p_{2,R_2}  & -\Delta \bm p_{2,R_2}
\label{eq:x1_and_x2}
\end{array}\right].
\end{align}
Note that $\bm X_1$ and $\bm X_2$ are each of size $3\times6$ with $\Delta \bm p_{i,j} = \bm p_i -\bm p_j $ and each $\bm p_i\in\mathbb{R}^2$ denoting a length-2 row vector encoding the $(x,y)$-coordinates of a coordinate-labeling point.
The $5\times8$ rigidity matrix of the composite system $R_1\ast R_2$ is given by
\begin{align}
\bm X_1\ast \bm X_2=\left[
\begin{array}{cccc}
 \Delta \bm p_{1,2} &  \Delta \bm p_{2,1} & \bm 0 &\bm 0\\
\Delta \bm p_{1,R_1}& \bm 0 & \Delta \bm p_{R_1,1} & \bm 0\\
\bm 0 &\Delta \bm p_{2,R_1} & \Delta \bm p_{R_1,2} &\bm 0\\
\Delta \bm p_{1,R_2} & \bm 0 &\bm 0 &\Delta \bm p_{R_2,1} \\
\bm 0 &\Delta \bm p_{2,R_2}& \bm 0 & \Delta \bm p_{R_2,2}
\label{rm_simple_2comp}
\end{array}\right],
\end{align}
where the first row derives from $R_1\cap R_2$, the second and third rows from $ R_1$, and the fourth and fifth rows from $R_2$.
We can group these elements into blocks such that the diagonal blocks are 
\begin{equation}
\left \{ \left[ \begin{array}{cc} \Delta \bm p_{1,2} & -\Delta \bm p_{1,2}\end{array}\right], \left[ \begin{array}{c} \Delta \bm p_{R_1,1}\\ \Delta \bm p_{R_1,2} \end{array}\right], \left[ \begin{array}{c} \Delta \bm p_{R_2,1}\\ \Delta \bm p_{R_2,2}\end{array}\right]\right \},
\end{equation}
each of which has full row rank (i.e., rank $1$, $2$, and  $2$, respectively), because $\bm p_{R_1}$ and $\bm p_{R_2}$ are each individually noncollinear with $\bm p_1$ and $\bm p_2$, by construction of the coordinate labelings. Because each diagonal block has full row rank, the block triangular matrix also has full row rank. Therefore, $\text{rank}(\bm X_1\ast \bm X_2) = 5$, $\text{dim}(\text{null}(\bm X_1\ast \bm X_2)) = 8-5= 3$, and the composition is rigid---that is, the minimum number of degrees of freedom for a rigidity matrix of a two-dimensional system is $3$. 

In the case that $r_1=1$ (a rigid component that is a single rod), $R_1$ only requires two points to specify its rigid motions, and so we choose these to be $\bm p_1$ and $\bm p_2$, giving the $3\times6$ rigidity matrix
\begin{align}
\bm X_1\ast \bm X_2=\left[
\begin{array}{ccc}
\Delta \bm p_{1,2} & -\Delta \bm p_{1,2}& \bm 0 \\
\Delta \bm p_{1,R_2} & \bm 0 & -\Delta \bm p_{1,R_2} \\
\bm 0 &\Delta \bm p_{2,R_2} & -\Delta \bm p_{2,R_2}
\label{rm_simple_2comp_2}
\end{array}\right],
\end{align}
which trivially has full row rank and thus a right nullspace dimension of $3$. Because individual distinct straight rods cannot intersect at more than one point, $r_1$ and $r_2$ cannot simultaneously both be one, so the two cases complete the proof.
\end{proof}
\end{theorem}
\medskip

The necessity of two contacts in this latter scenario (where $r_1=1$) begs the following (rather obvious) lemma, which differentiates rigidity percolation from contact percolation:
\begin{lemma}
\label{degree1}
Every rod in a rigid 2D rod-hinge system (with $r>1$) must have at least two contacts. 
\begin{proof} See Appendix \ref{app:motif_proofA}.
\end{proof}
\end{lemma}
\medskip


\begin{theorem}[Motif 2D3]\label{thm:2D3} 
The composition of three rigid components $R_1$, $R_2$, and $R_3$ intersecting pairwise in two dimensions at three or more points including $\bm p_1\in (R_1\cap R_3),$ $\bm p_2 \in (R_1\cap R_2)$, $\bm p_3 \in (R_2\cap R_3)$, is rigid. 
\begin{proof} See Appendix \ref{app:motif_proofB}.
\end{proof}
\end{theorem}
\medskip

\begin{theorem}[Motif 2D5] \label{thm:2D5} 
If five rigid components $\{R_{1},\dots,R_5\}$ intersect in two dimensions at six or more points such that $\bm p_1\in (R_1\cap R_3)$, $\bm p_2 
\in (R_2\cap R_3)$, $\bm p_3 \in (R_1\cap R_4)$, $\bm p_4 \in (R_2 \cap R_4)$, $\bm p_5 \in (R_1\cap R_5),$ $\bm p_6\in(R_2\cap R_5)$, then their composition is rigid---except in the degenerate case with $\{\bm p_1,\bm p_3, \bm p_5\}$ and $\{\bm p_2, \bm p_4,\bm p_6\}$ each collinear (rod-sharing) sets and the vectors $\Delta \bm p_{1,2}$, $\Delta \bm p_{3,4}$, $\Delta \bm p_{5,6}$ are mutually parallel (which occurs with probability $0$).
\begin{proof} See Appendix \ref{app:motif_proofC}.
\end{proof}
\end{theorem}
\medskip

\subsection{Rigid Graph Compression (RGC) in 2 Dimensions}\label{sec:rgc}
We now proceed to use the primitive rigid motifs described above to identify large-scale rigid components agglomerated from rigid components identified at smaller scales, starting from the microscopic scale of primitive rigid motifs acting on individual rods. In so doing, it is convenient to work in terms of  \emph{rod contact networks} in which each node represents a rigid component and edges indicate which components intersect with one another. Importantly, this network construction contrasts the \emph{constraint graphs} described earlier (in which nodes represented coordinate labelings and edges represent rigidity constraints). 

\begin{figure}
\begin{center}
\includegraphics[width=.65\linewidth]{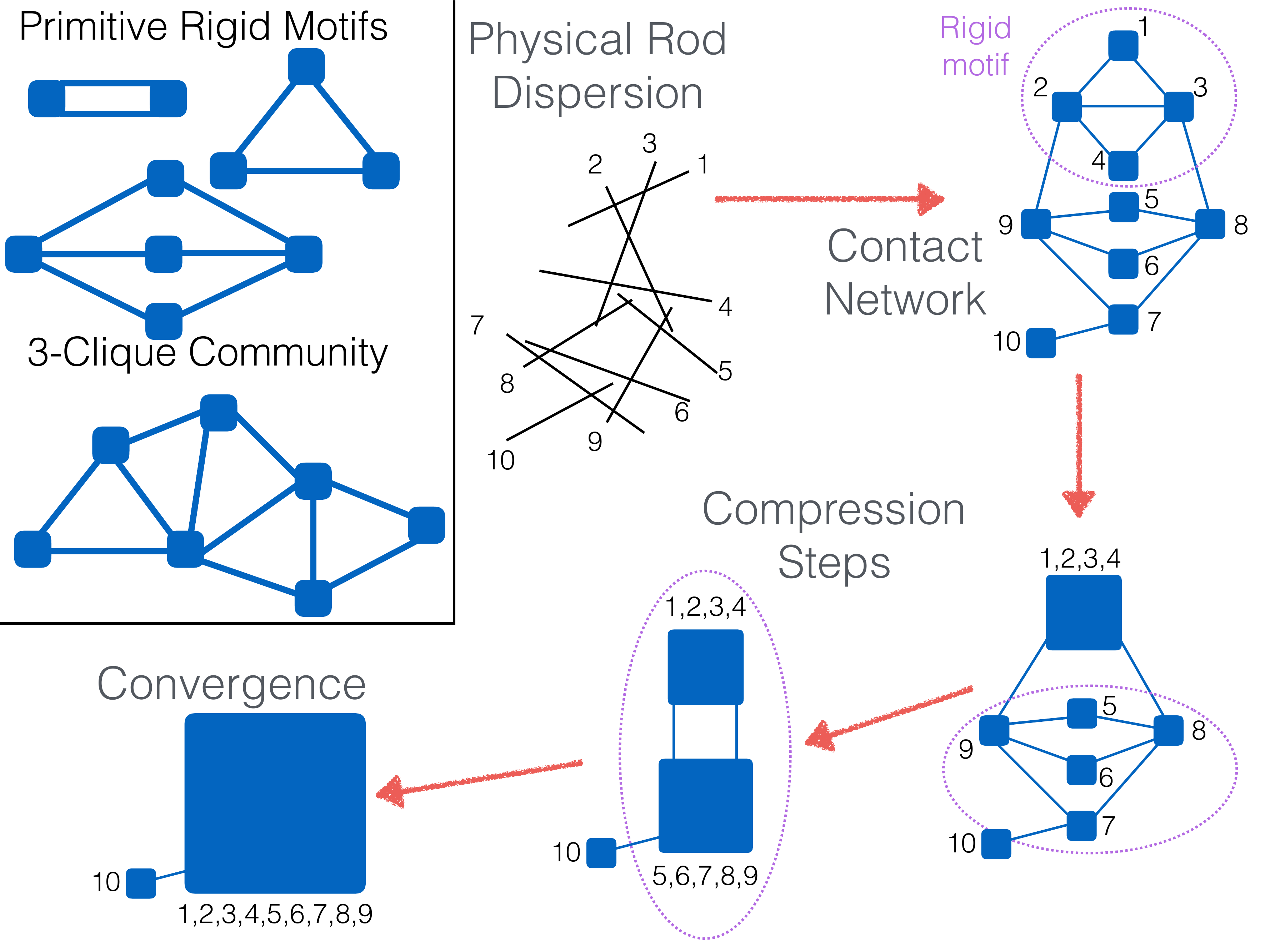}
\caption{
{\bf Graph compression of rod-hinge systems using rigid motifs} 
Using a 10-component rod-hinge system as an example, we describe \emph{2D-RGC-5} (Algorithm~\ref{alg:2DRGC}), which iteratively compresses 2- and 5-component primitive rigid motifs, as well as 3-clique communities (see top left inset for contact network representations of these motifs).
In the first step, the physical rod dispersion is transformed into a rod contact network. This contact network contains both a 3-clique community (nodes 1-4) and a 5-component motif (5-9). In two steps, each of these motifs are compressed into a single compound node. These two composite nodes are connect by two edges, which is the 2-component primitive rigid motif and is then compressed in the final step, giving one compound node representing rods 1-9 connected to another node representing rod 10. Stopping in the absence of any other primitive rigid motifs, RGC thus identifies two rigid components within the candidate rod-hinge system.
}
\label{fig:rgc}
\end{center}
\end{figure}

Before describing our graph compression algorithm, we note that we employ an additional method to enhance the algorithm's computational efficiency. Rather than identifying instances of Motif 2D3 directly, we make use of an available fast algorithm  \cite{cpm,networkx} for identifying \emph{k-clique communities}, which are sets of $k$-cliques (complete subgraphs on $k$ nodes), joined pairwise at $k-1$ points. In particular, any $3$-clique community (see Fig~\ref{fig:rgc}) is necessarily a composite rigid motif in 2D, by repeated application of Motifs 2D2 and 2D3. Furthermore, every instance of Motif 2D3 is a member of a $3$-clique communities (at that spatial scale), although the same cannot be said for Motif 2D2. 

We now introduce the graph-compression algorithm that leverages the motifs described above. The RGC algorithm involves initialization (Step 1), followed by iterative identification and compression of rigid motifs (Steps 2 \& 3):
(1) given a set of interacting particles (i.e.\ a rod-hinge system), construct a contact network of rods represented as nodes and contacts between rods  represented as edges; (2) identify rigid motifs in the contact network; (3) compress each rigid motif instance into a single node, yielding a reduced set of rigid components and an updated contact network; (4) return to Step 2.

The format of this algorithm can be adapted to include any number of motifs in any number of dimensions. In the experiments of the next section, we incorporate the three primitive rigid motifs of Sec.~\ref{sec:motifs} along with 3-clique communities for 2D rod-hinge systems in \emph{2D-RGC-5}, denoting that the algorithm utilizes all primitive rigid motifs that include up to 5 rigid bodies. We compare results from \emph{2D-RGC-5} with the same methodology using only the 2-component motif and 3-clique communities, which we label \emph{2D-RGC-3} to denote that it only utilizes the primitive rigid motifs that include up to 3 rigid bodies. Pseudocode for \emph{2D-RGC-5} is provided in Algorithm~\ref{alg:2DRGC}. We note that, after initial creation of the rod contact network representation, we identify instances of Motif 2D5 and of 3-clique communities. Then, we compress each motif instance, condensing its aggregating nodes into one surviving node, and rewiring all of the out-edges from the aggregating nodes to this surviving node; importantly, this can create edges with weights of 2 when some of these out-edges are repeated. In this reduced network, we again identify instances of Motif 2D5 and 3-clique communities, as well as of Motif 2D2, before compressing again, iterating until none of the motifs are present in the network (see Fig.~\ref{fig:rgc} for an example).

\begin{algorithmic}[ht]
\begin{algorithm}[t]
\caption{\bf:   Rigid Graph Compression (2D-RGC-5)}\label{alg:2DRGC}
\State{Generate Contact Graph $G(V,E)$ from $\{R\}$}
\While{$\exists$ $3$-clique communities OR $2$-component motifs OR $5$-component motifs $\in G$}
	\State{Identify $3$-clique communities in $G$}
	\For{each 3-clique community in $G$}
		\State{Compress($G$,\{nodes in $3$-clique community\}, \{edges in $3$-clique community\})}
	\EndFor
	\State{Identify $2$-component motifs in $G$}
	\For{each $2$-component motif in $G$}
		\State{Compress($G$,\{nodes in $2$-component motif\}, \{edges in $2$-component motif\})}
	\EndFor
	\State{Identify $5$-component motifs in $G$}
	\For{each $5$-component motif in $G$}
		\State{Compress($G$,\{nodes in $5$-component motif\}, \{edges in $5$-component motif\})}
	\EndFor
	\State{Identify $3$-clique communities, $2$-component motifs, $5$-component motifs in $G$}
\EndWhile\\
\Procedure{Compress}{$G,\{nodes\},\{edges\}}$
  \State Rewire all out-edges within $\{edges\}$ to a node $x\in nodes$, assign a weight of two to any out-edges that are rewired $\ge 2$ times
  \State Delete all nodes in $\{nodes\}$ except $x$.\\
\EndProcedure
\end{algorithm}
\end{algorithmic}

Finally, we note that while we could choose to order the three motif compression steps of Algorithm~\ref{alg:2DRGC} in $3!=6$ possible ways, this order does not affect our analysis. In the case that rigid motifs do not intersect, any motif compression order trivially gives the same result; but when two or more rigid motifs intersect (their contact network representations share at least one node), this conclusion is slightly more nuanced. In Appendix~\ref{app:order}, we argue that any motif compression order will yield the same results when Algorithm~\ref{alg:2DRGC} is subjected to intersecting motifs. 


\section{Rigidity of 2D Rod-Hinge Systems: Numerical Experiments}\label{sec:results}
As has previously been found using the pebble game and dynamical simulation \cite{pebgame2drods2,stiffpolymer}, 2D rod-hinge systems undergo a rigidity percolation transition with respect to rod density that is markedly above the contact percolation transition. Here we demonstrate that RGC accurately identifies this transition. We construct the rod-hinge systems using the same method as \cite{pebgame2drods2}: unit-length rods are placed with uniformly random position and orientation in a rectangular domain. This domain is divided into an $L\times L$ square region, with $1\times L$ `buffer regions' on the left and right sides along one dimension, to eliminate bias in the rod density near these boundaries. We effectively place a large, length $L$ (infinitesimally thin) rod along the boundary between each buffer region with the interior square domain, and define rigidity percolation by the presence of a spanning rigid component containing both of these boundaries (see Fig.~\ref{fig:rods1}). We use periodic boundary conditions in the other dimension.

\begin{figure}[t]
\begin{center}
\includegraphics[width=\linewidth]{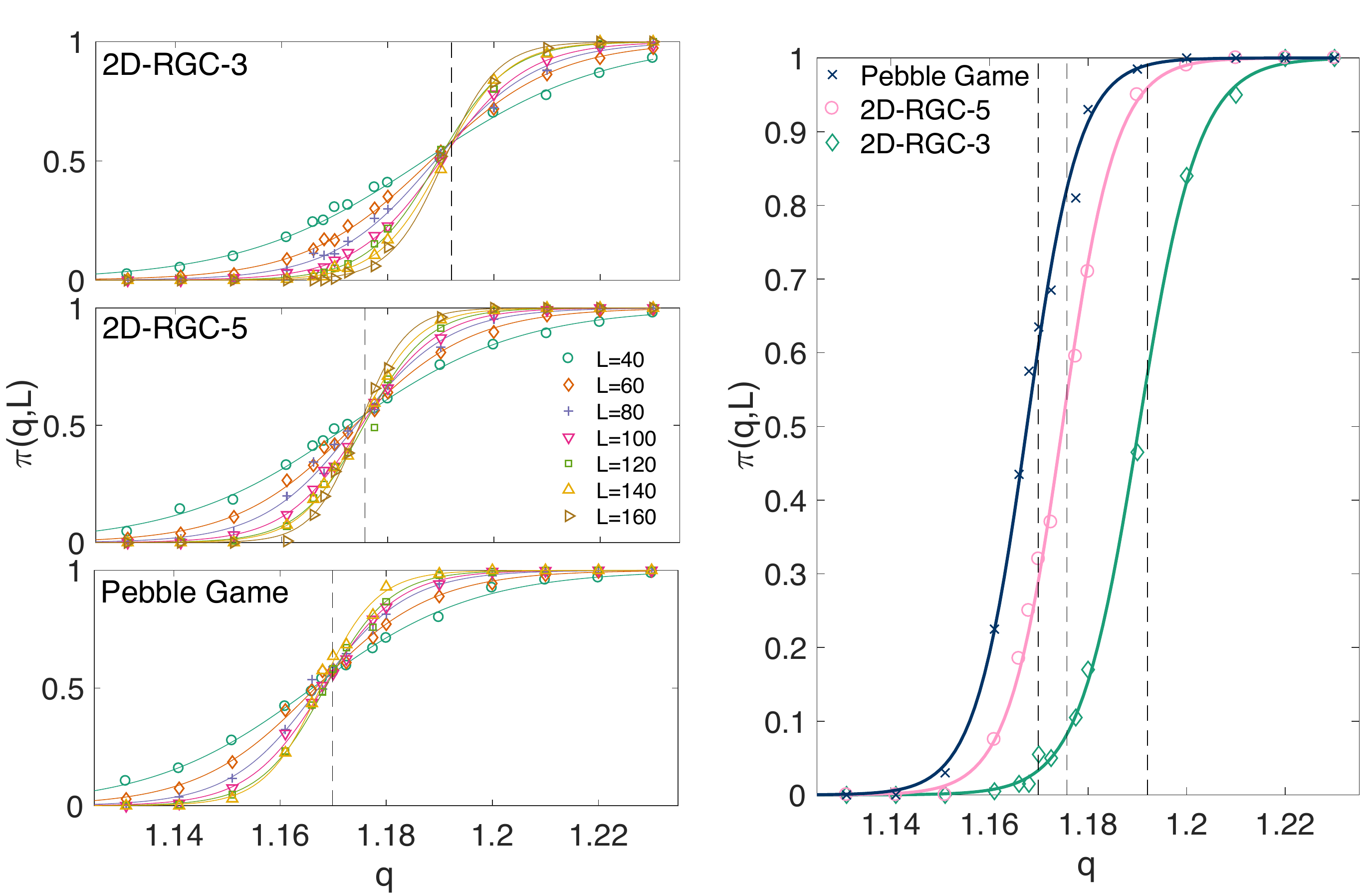}
\caption{{\bf Comparison of 2D-RGC-3, 2D-RGC-5, and  pebble-game algorithms for 2D rigidity percolation.} 
Left: For all three rigidity-detection algorithms, there is a phase transition in $\pi(q,L)$ that becomes sharper with increasing $L$---an extrapolation algorithm is used to estimate rigidity percolation thresholds (vertical dashed lines) from these individual curves. The transitions identified using the RGC algorithms approximate that of the pebble game, with that of the \emph{2D-RGC-5} being the closer approximation. Incorporation of yet more rigid motifs would further increase the accuracy of this approximation. Right: Rigidity percolation transitions for each of the three algorithms are displayed for a large domain size, $L=140$. 
}
\label{fig:plotall}
\end{center}
\end{figure}

We generated rod dispersions in domains of size $L=40$, $60$, $80$, $100$, $120$, $140$ and $160$. We considered 15 different rod densities $q={r}/({q_cL^2})$, where $q_c\doteq 5.71$ is the contact-percolation threshold for 2D rod-hinge systems \cite{statgeoclassic}, centered about the rigidity percolation threshold estimate from \cite{pebgame2drods2}. In each simulated dispersion, we checked for the presence of a spanning rigid component using each of three  algorithms: \emph{2D-RGC-3} and \emph{2D-RGC-5} (described in Sec.~\ref{sec:rgc}); and the pebble game (Sec.~\ref{subsec:laman}). We used between $150$ and $1100$ simulation trials at each $(q,L)$ pair to approximate the probability of rigidity percolation $\pi(q,L)$ across this parameter space (see Fig.~\ref{fig:plotall}).

\begin{figure}[t]
\begin{center}
\includegraphics[width=\linewidth]{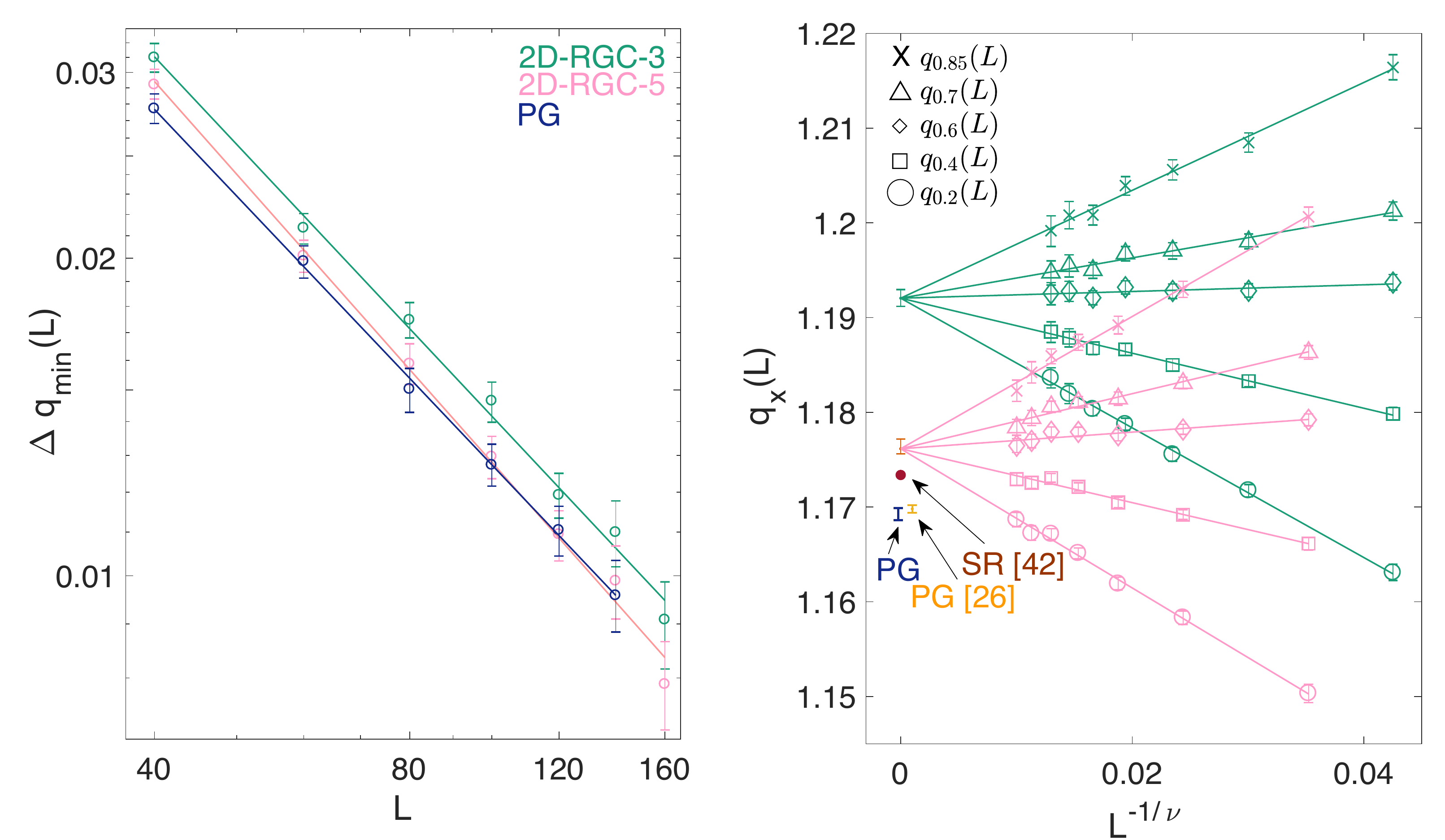}
\caption{{\bf Estimation of correlation length exponent and rigidity percolation threshold for RGC and pebble game algorithms.} Left: Using each rigidity characterization algorithm, we use the relation $\Delta q_\mathrm{min}(L)\sim L^{-1/\nu}$ to estimate $\nu$. 
Right: An extrapolation scheme is used to estimate $q_\mathrm{min}$ using each of the three rigidity detection algorithms. For comparison, we display the rigidity percolation threshold found in \cite{pebgame2drods2} using the pebble game (PG [26]), in our own pebble game calculations (PG), and in \cite{stiffpolymer} using spring relaxation (SR). 
}
\label{fig:correlation}
\end{center}
\end{figure}

In order to accurately estimate the rigidity percolation threshold $q_\mathrm{min}$ and correlation length exponent $\nu$ (associated with the divergence of the correlation length about $q_\mathrm{min}$) corresponding to each algorithm, we first assume a data collapse in accordance with classical percolation theory \cite{pebgame2drods2,perctheory}: 
\begin{align} 
\pi(q-q_\mathrm{min},L)=\phi([q-q_\mathrm{min}]L^{1/\nu}),
\label{eq:datacollapse}
 \end{align}
from which we find $\frac{d\pi}{dq}=L^{1/\nu}\phi'([q-q_\mathrm{min}]L^{1/\nu})$ for $q\to q_\mathrm{min}$. As in \cite{pebgame2drods2}, we invert the scaling of $d\pi/dq$ with $L^{1/\nu}$, finding that 
\begin{align} \Delta q_\mathrm{min}(L):=\langle \sqrt{(q_{est}-q_{av})^2}\rangle
\end{align}
scales as $L^{-1/\nu}$, where $q_{est}$ is the density at which a spanning cluster first appears for a particular set of simulations at a given $L$ and $q_{av}$ is the average of these simulations (angular brackets denote averages). For each algorithm and domain size $L$, we estimate $\Delta q_\mathrm{min}(L)$ by fitting a cumulative logistic distribution $F(q;\mu,s)={1}/({1+e^{-\frac{q-\mu}{s}}})$ to the set of values $\pi(q,L)$, and then setting $\Delta q_\mathrm{min}(L)$ equal to the standard deviation ${s\pi}/{\sqrt{3}}$ of this distribution. We fit $\log \Delta q_\mathrm{min}(L)$ versus $\log L$ (left panel of Fig.~\ref{fig:correlation}) via least squares minimization to estimate $\nu$, using a simple case resampling method to simultaneously determine confidence intervals for every $\Delta q_\mathrm{min}(L)$ estimation. From each set of $\Delta q_\mathrm{min}(L)$ samples, we estimate a fit for $\nu$ and use the collection of these fits to calculate the corresponding confidence intervals. We thereby obtain $\nu\doteq 1.1682,1.1025,1.1812$ for \emph{2D-RGC-3}, \emph{2D-RGC-5}, and the pebble game, respectively, with corresponding 95\% confidence intervals of $[1.1137,1.2546]$, $[1.0474,1.1670]$, and $[1.1036,1.2340]$. While the correlation length exponent estimated by \emph{2D-RGC-5} is comparatively low, we note the confidence intervals from the three methods overlap. 

\begin{figure}[t]
\centering
	\includegraphics[width=.8\linewidth]{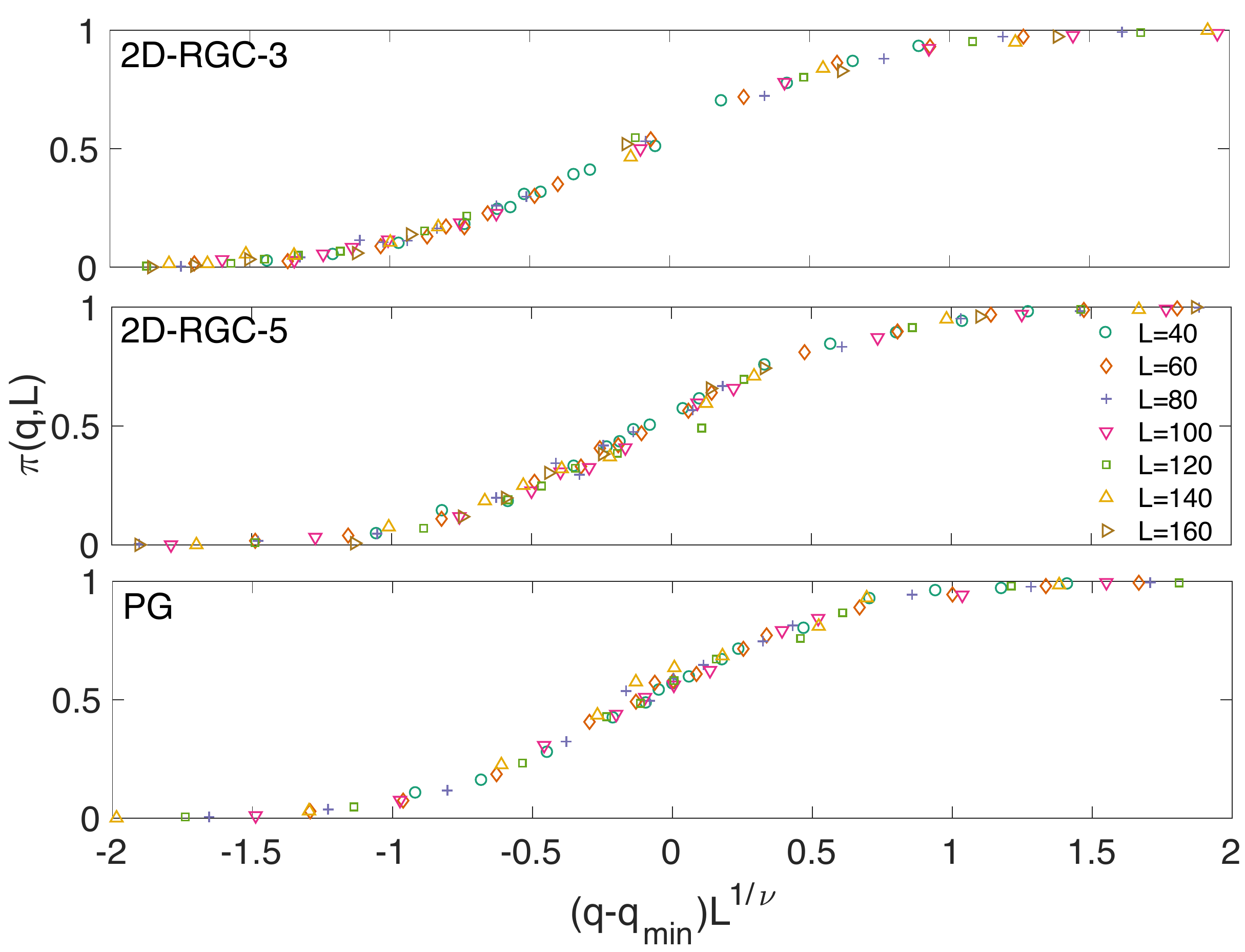}
	\caption{{\bf Demonstration of data collapse.} Using the identified values of $\nu$ and $q_\mathrm{min}$ for each rigidity detection algorithm, we find the data collapse assumption (according to Eq.~\ref{eq:datacollapse}) to be quite sound.
	}
	\label{fig:datacollapse}
\end{figure}

Having estimated $\nu$, we seek to now estimate  $q_\mathrm{min}$. To derive a scaling law, we expand $\pi(q,L)$ around $q=q_\mathrm{min}$ in Eq.~\ref{eq:datacollapse} and invert, deriving the condition: 
\begin{align} 
q_x(L)=\mathrm{(constant)}\cdot L^{-1/\nu} +q_\mathrm{min}\,, 
\end{align} 
where $q_x(L)$ is a probability distribution such that $\pi(q_x(L),L)=x$ for some $x\in[0,1]$ \cite{pebgame2drods2,perctheory}. We use this equation to extrapolate the $q_x(L)$ values as $L\to\infty$ to predict $q_\mathrm{min}$ for each algorithm as follows. First, we find $q_x(L)$ via inverse prediction from the corresponding cumulative distribution $F(q;\mu,s)$ for $x=0.2,0.4,0.6,0.7,0.85$. Then, we fit each set of $\log q_x(L)$ values against $-\nu^{-1} \log L$, with the constraint that each of these fits must coincide at the intercept with the $q$ axis (see right panel of Fig.~\ref{fig:correlation}). We estimate the intercepts to be $q_\mathrm{min}\doteq 1.1920,$ $1.1757,$ and $1.1692$ for \emph{2D-RGC-3}, \emph{2D-RGC-5}, and the pebble game, respectively, with corresponding 95\% confidence intervals of [1.1912, 1.1929], [1.1756, 1.1767], and [1.1686, 1.1698]. Taking the pebble game estimate to be the true threshold, we find the relative errors for the \emph{2D-RGC-3} and \emph{2D-RGC-5} estimates to be $1.9\%$ and $0.6\%$, respectively. Finally, having identified $q_\mathrm{min}$ and $\nu$ for each rigidity-detection algorithm, we confirm that the rescaling data collapse $\pi(q,L)=\phi([q-q_\mathrm{min}]L^{1/\nu})$ assumption is quite accurate (see Fig.~\ref{fig:datacollapse}).

\begin{figure}[t]
\centering
\includegraphics[width=.5\linewidth]{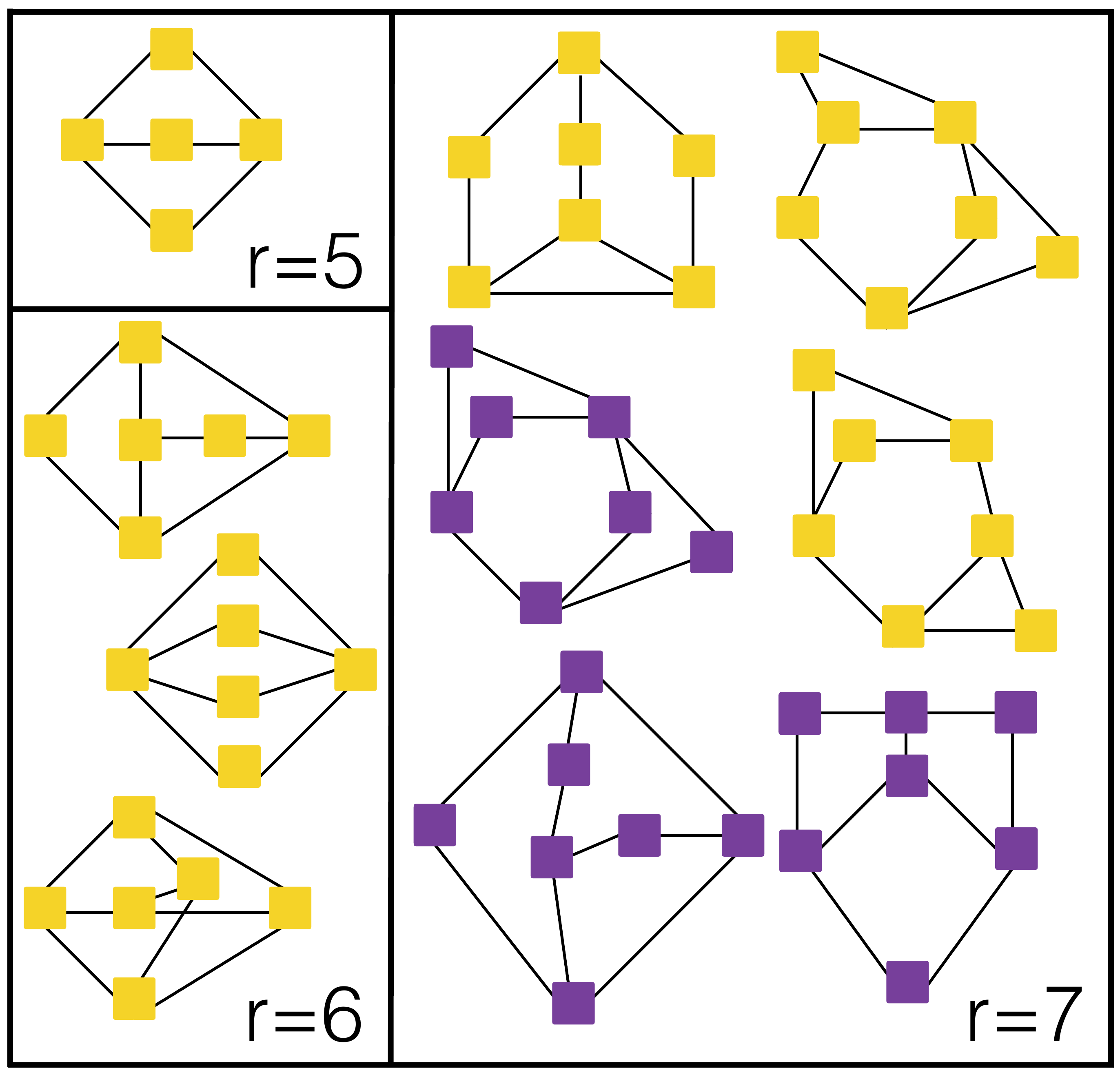}
\caption{
{\bf Rigid motifs not identified by 2D-RGC-3 and/or 2D-RGC-5.} Exhaustive search of rod contact networks containing up to seven rods reveals seven rigid motifs incorrectly identified as floppy by 2D-RGC-3 only (yellow), and three other rigid motifs incorrectly identified as floppy by both 2D-RGC-3 and 2D-RGC-5 (purple). These latter motifs---which are classified as rigid via the pebble game---could potentially be incorporated into a 2D-RGC-7 algorithm.}
\label{fig:graphlets}
\end{figure}

To better understand the discrepancies between the rigidity percolation transition as predicted by RGC and the pebble game, we apply each algorithm to graphlets---small connected nonisomorphic graphs---in the rod contact network representation. We limit our search to graphlets with minimal degree two (on account of Lemma~\ref{degree1}). We confirm that \emph{2D-RGC-3} perfectly characterizes rigidity for all graphlets of $r\le 4$ components, but of course fails to detect that the $r=5$ Motif 2D5 is rigid, and thus misses a number of $r>5$ cases as well (the yellow motifs in Fig.~\ref{fig:graphlets}). By accounting for the rigidity of Motif 2D5, \emph{2D-RGC-5} is accurate for all graphlets with $r\le 6$, but misses three $r=7$ cases (purple motifs in Fig.~\ref{fig:graphlets}). By incorporating these primitive rigid motifs with $7$ components, one might develop a \emph{2D-RGC-7} algorithm and estimate $q_\mathrm{min}$ with even higher accuracy relative to the two RGC versions used in this study.

\section{Discussion}
We have developed an efficient computational approach for accurately characterizing the rigidity of 2D rod-hinge systems in a manner that can be easily extended to three dimensions, for which the problem remains open. We identified a small set of primitive rigid motifs (see Fig.~\ref{fig:axioms}) as building blocks of larger aggregate rigid components and developed an algorithm, Rigid Graph Compression (RGC), to collapse {a rod contact network accordingly, thus building up to spatially-extended rigid components. We demonstrated the effectiveness of our approach by characterizing rigidity percolation in 2D random rod dispersions, comparing the RGC results to those of the pebble game (which is exact in 2D but does not naturally extend to higher dimensions). 
Denoting the RGC algorithm using primitive rigid motifs up to $y$ interacting rods by \emph{2D-RGC-y}, we find the accuracy increases asymptotically with $y$: \emph{2D-RGC-3} approximates the rigidity percolation threshold to within 2\% relative error, and \emph{2D-RGC-5} approximates the threshold with $\doteq$ 0.6\% relative error.

Given these encouraging results, future work will develop and apply analogous primitive rigid motifs to analyze 3D rod-hinge systems, which have a more expansive range of applications for modeling real physical systems. In particular, we model nano-rod composites using rod-hinge systems and hypothesize that the experimentally observed mechanical percolation phase transitions in composites can be effectively studied through rigidity percolation in rod-hinge systems. As there are no sufficient existing rigidity characterization algorithms for comparison, the task of determining accuracy is less straightforward in 3D systems---thus, the present 2D study was essential to validate the feasibility of the RGC approach in general. We note two interesting differences between 2D and 3D rod dispersions, highlighting complications that must be overcome pursuing this direction. First, infinitesimally thin rods do not generically intersect in 3D, requiring a second independent variable (in addition to rod density) to inform meaningful rod-contact data with finite aspect ratio. Second, rods and larger-scale rigid components both have three degrees of freedom in 2D, and so the 2D rigidity motifs we develop are independent of this scale; however, in 3D, if one does not account for rotation around the axis of symmetry, rods have five degrees of freedom whereas multiple-rod rigid components have six, and therefore the 3D rigid motifs must distinguish these two classes.

\appendix

\section{Proof of Lemma~\ref{degree1}\label{app:motif_proofA}}
Suppose that a single rod $R_1$ has exactly one contact $\bm p_1$ with a rigid component $R_2$. Only two points are required to specify the rigid motions of $R_1$; we choose these to be $\bm p_1$ and $\bm p_{R_1}$. For ${R_2}$, we choose the coordinate labeling $\{\bm p_1, \bm p_{R_{2a}}, \bm p_{R_{2b}}\}$. The corresponding $4\times8$ rigidity matrix is
\begin{align}
\bm X_1\ast \bm X_2=\left[
\begin{array}{cccc}
\Delta \bm p_{1,R_1} & -\Delta \bm p_{1,R_1} & \bm 0& \bm 0 \\
\Delta \bm p_{1,R_{2a}} & \bm 0 & -\Delta \bm p_{1,R_{2a}} & \bm 0 \\
\Delta \bm p_{1,R_{2b}} &\bm 0 &\bm 0 & -\Delta \bm p_{1,R_{2b}}\\
\bm 0 &\bm 0 & \Delta \bm p_{R_{2a},R_{2b}} & -\Delta \bm p_{R_{2a},R_{2b}}\\
\end{array}\right],
\end{align}
which trivially cannot have rank $>4$ and thus its right nullspace dimension is at least $4$, implying the composition is not rigid.
\endproof

\section{Proof of Theorem~\ref{thm:2D3}\label{app:motif_proofB}}
First, suppose $r_1>1$, $r_2>1$, and $r_3>1.$ As in the proof of Thm.~\ref{axiom1_simple}, we choose a minimal coordinate labeling for each rigid component, each of which includes two intersection points and one additional point, giving the sets of labelings: $\{ \bm p_{R_1}, \bm p_1, \bm p_2\},$ $\{ \bm p_{R_2}, \bm p_2, \bm p_3\},$ and $\{ \bm p_{R_3}, \bm p_3, \bm p_1\}$.
The composition's $9\times12$ rigidity matrix is larger in this scenario:
\begin{align}
\bm X_1\ast \bm X_2\ast \bm X_3=\left[
\begin{array}{cccccc}
\Delta \bm p_{1,2} & -\Delta \bm p_{1,2} & \bm 0 & \bm 0& \bm 0& \bm 0 \\
\Delta \bm p_{1,R_1} & \bm 0 & \bm 0 & -\Delta \bm p_{1,R_1} & \bm 0 & \bm 0 \\
\bm 0 &\Delta \bm p_{2,R_1} & \bm 0 & -\Delta \bm p_{2,R_1} & \bm 0 & \bm 0\\
\bm 0 & \Delta \bm p_{2,3} & -\Delta \bm p_{2,3} & \bm 0 & \bm 0 & \bm 0\\
\bm 0 & \Delta \bm p_{2,R_2} & \bm 0 & \bm 0 & -\Delta \bm p_{2,R_2} & \bm 0\\
\bm 0 & \bm 0 & \Delta \bm p_{3,R_2} & \bm 0  & -\Delta \bm p_{3,R_2} & \bm 0\\
\Delta \bm p_{1,3} & \bm 0 & -\Delta \bm p_{1,3} & \bm 0 & \bm 0 & \bm 0\\
\Delta \bm p_{1,R_3} & \bm 0 & \bm 0 & \bm 0 & \bm 0 & -\Delta \bm p_{1,R_3}\\
\bm 0 & \bm 0 & \Delta \bm p_{3,R_3} & \bm 0 & \bm 0 & -\Delta \bm p_{3,R_3}
\label{rm_simple_3comp_1}
\end{array}\right].
\end{align}
The first three rows derive from $R_1$, the second three from $R_2,$ and the third from $R_3$.
Row permutation of $\bm X_1\ast \bm X_2\ast \bm X_3$ gives
\begin{align}
\left[
\begin{array}{cccccc}
\Delta \bm p_{1,2} & -\Delta \bm p_{1,2} & \bm 0 & \bm 0& \bm 0& \bm 0 \\
\Delta \bm p_{1,3} & \bm 0 & -\Delta \bm p_{1,3} & \bm 0 & \bm 0 & \bm 0\\
\bm 0 & \Delta \bm p_{2,3} & -\Delta \bm p_{2,3} & \bm 0 & \bm 0 & \bm 0\\
\Delta \bm p_{1,R_1} & \bm 0 & \bm 0 &-\Delta \bm p_{1,R_1} & \bm 0 & \bm 0 \\
\bm 0 &\Delta \bm p_{2,R_1} & \bm 0 & -\Delta \bm p_{2,R_1} & \bm 0 & \bm 0\\
\bm 0 & \Delta \bm p_{2,R_2} & \bm 0 & \bm 0 & -\Delta \bm p_{2,R_2} & \bm 0\\
\bm 0 & \bm 0 & \Delta \bm p_{3,R_2} & \bm 0  & -\Delta \bm p_{3,R_2} & \bm 0\\
\Delta \bm p_{1,R_3} & \bm 0 & \bm 0 & \bm 0 & \bm 0 & -\Delta \bm p_{1,R_3}\\
\bm 0 & \bm 0 & \Delta \bm p_{3,R_3} & \bm 0 & \bm 0 & -\Delta \bm p_{3,R_3}
\end{array}\right].
\end{align}
The same block diagonal argument from the proof in Thm~\ref{axiom1_simple} may be applied to show that this matrix has full row rank and thus right nullspace dimension $3$. Explicitly, the diagonal blocks 
\begin{equation}
\left \{ \left[\begin{array}{cc}
\Delta \bm p_{1,2} & -\Delta \bm p_{1,2} \end{array}\right], \left[\begin{array}{c} -\Delta \bm p_{1,3}\\  -\Delta \bm p_{2,3} \end{array}\right], \left[\begin{array}{c} -\Delta \bm p_{1,R_1}\\  -\Delta \bm p_{2,R_1} \end{array}\right], \left[\begin{array}{c} -\Delta \bm p_{2,R_2}\\  -\Delta \bm p_{3,R_2} \end{array}\right], \left[\begin{array}{c} -\Delta \bm p_{1,R_3}\\ -\Delta \bm p_{3,R_3} \end{array}\right] \right \}
\end{equation}
are each full row rank because the three points in a coordinate labeling are necessarily noncollinear. Therefore, the matrix is of full row rank (9) and the dimension of the right nullspace is $12-9=3$, which is again the minimum for all two-dimensional network rigidity matrices.

If one or more of the individual rigid components are single rods, then for each of these we may drop one of $\bm p_{R_1}$, $\bm p_{R_2}$, or $\bm p_{R_3}$, as well as the two corresponding constraints, giving no net change in the dimension of the right nullspace. Therefore, the composition is rigid.
\endproof

\section{Proof of Theorem~\ref{thm:2D5}\label{app:motif_proofC}}
In the first case, we assume that the intersection points contained in $R_1$ ($\bm p_1, \bm p_3,$ and $\bm p_5$) are noncollinear  (they do not all lie along the same rod) as are those contained in $R_2$ ($\bm p_2, \bm p_4,$ and $\bm p_6$). We also assume that each rigid body is a compound object of multiple rods (i.e., $r_1>1,...,r_5>1$). Then, we choose as coordinate labelings: $\{\bm p_1, \bm p_3\,\bm p_5\}$ for $R_1$,  $\{\bm p_2, \bm p_4\,\bm p_6\}$ for $R_2$, $\{\bm p_1, \bm p_2, \bm p_{R_3}\}$ for $R_3$, $\{\bm p_3, \bm p_4, \bm p_{R_4}\}$ for $R_4$, and $\{\bm p_5, \bm p_6, \bm p_{R_5}\}$ for $R_5$. These labelings inform $5\cdot 3=15$ constraints, given in the rigidity matrix:

\begin{multline} \bm X_1\ast... \ast \bm X_5= \\
\scriptsize{ \left[
\begin{array}{ccccccccccc}
\Delta \bm p_{1,2}&- \Delta \bm p_{1,2} &\bm 0 & \bm 0 & \bm 0& \bm 0& \bm 0& \bm 0& \bm 0 \\
\Delta \bm p_{1,3} & \bm 0&-\Delta \bm p_{1,3}  & \bm 0 & \bm 0 & \bm 0& \bm 0 &\bm 0 &\bm 0 \\
\bm 0& \Delta \bm p_{2,4}&\bm 0 & -\Delta \bm p_{2,4} &  \bm 0 & \bm 0& \bm 0 & \bm 0& \bm 0 \\
\bm 0& \bm 0&\Delta \bm p_{3,4} & -\Delta \bm p_{3,4} & \bm 0&  \bm 0& \bm 0 & \bm 0& \bm 0 \\
\Delta \bm p_{1,5} & \bm 0&\bm 0& \bm 0 & -\Delta \bm p_{1,5} & \bm 0& \bm 0 &\bm 0 &\bm 0 \\
\bm 0& \bm 0&\Delta \bm p_{3,5}  & \bm 0 & -\Delta \bm p_{3,5}   & \bm 0& \bm 0 &\bm 0 &\bm 0 \\
\bm 0& \Delta \bm p_{2,6}&\bm 0 & \bm 0 & \bm 0& -\Delta \bm p_{2,6}& \bm 0 & \bm 0& \bm 0 \\
\bm 0& \bm 0&\bm 0 & \Delta \bm p_{4,6} & \bm 0& -\Delta \bm p_{4,6}&  \bm 0 & \bm 0& \bm 0 \\
\bm 0& \bm 0&\bm 0 & \bm 0 & \Delta \bm p_{5,6}& -\Delta \bm p_{5,6}& \bm 0 & \bm 0& \bm 0 \\
\Delta \bm p_{1,R_3}&\bm 0&\bm 0 & \bm 0 & \bm 0& \bm 0 &-\Delta \bm p_{1,R_3} & \bm 0& \bm 0 \\
\bm 0& \Delta \bm p_{2,R_3}&\bm 0 & \bm 0 & \bm 0& \bm 0 & -\Delta \bm p_{2,R_3} & \bm 0& \bm 0 \\
\bm 0& \bm 0&\Delta \bm p_{3,R_4} & \bm 0 & \bm 0& \bm0& \bm 0& -\Delta \bm p_{3,R_4}& \bm 0 \\
\bm 0& \bm 0&\bm 0 & \Delta \bm p_{4,R_4} & \bm 0&  \bm 0& \bm 0& -\Delta \bm p_{4,R_4}& \bm 0 \\
\bm 0& \bm 0&\bm 0 & \bm 0 & \Delta \bm p_{5,R_5}& 0& \bm 0 & \bm 0&  -\Delta \bm p_{5,R_5} \\
\bm 0& \bm 0&\bm 0 & \bm 0& \bm 0& \Delta \bm p_{6,R_5}& \bm 0 & \bm 0& -\Delta \bm p_{6,R_5}
\end{array}\right].}
\label{rm_noparallel}
\end{multline}
As in Appendix~\ref{app:motif_proofB}, we have arranged the rows for convenience in rank computation---rows $2,5,6$ derive from $R_1$; $3,7,8$ from $R_2$; $1,10,11$ from $R_3$; $4,12,13$ from $R_4$; and $9,14,15$ from $R_5$. This matrix may be partitioned into the diagonal blocks:
$$\bm A=\left[ \begin{array}{cc} \Delta \bm p_{1,2}&- \Delta \bm p_{1,2} \end{array}\right]\,,$$
$$
\footnotesize{ \bm B=\left[ \begin{array}{cccc}-\Delta\bm p_{1,3} & \bm 0 & \bm 0& \bm 0 \\
\bm 0& -\Delta \bm p_{2,4} & \bm 0 & \bm 0 \\
\Delta \bm p_{3,4} &-\Delta \bm p_{3,4} &\bm 0 &\bm  0 \\
\bm 0 & \bm 0 & -\Delta \bm p_{1,5}& \bm 0 \\
\Delta \bm p_{3,5} & \bm 0 & -\Delta \bm p_{3,5}& \bm 0\\
\bm 0& \bm 0 &\bm 0 & -\Delta \bm p_{2,6}\\
\bm 0& \Delta \bm p_{4,6} &\bm 0 & -\Delta \bm p_{4,6}\\
\bm 0& \bm 0 &\Delta \bm p_{5,6} & -\Delta \bm p_{5,6}
 \end{array}\right], \;\;
 \bm C= \left[\begin{array}{ccc}
-\Delta \bm p_{1,R_3} & \bm 0& \bm 0 \\
 -\Delta \bm p_{2,R_3} & \bm 0& \bm 0 \\
 \bm 0& -\Delta \bm p_{3,R_4}& \bm 0 \\
 \bm 0& -\Delta \bm p_{4,R_4}& \bm 0 \\
\bm 0& \bm 0&  -\Delta \bm p_{5,R_5} \\
 \bm 0& \bm 0& -\Delta \bm p_{6,R_5}
 \end{array}\right]}.
$$
Matrix $\bm A$ trivially has full row rank and matrix $\bm C$ has full rank due to the points in each coordinate labeling being noncollinear. To study $\bm B$, we can use a series of elementary row operations to eliminate the last six entries in row $B_{6,\bullet}$ and then place this modified $B_{6,\bullet}$ between $B_{1,\bullet}$ and $B_{2,\bullet}$, giving that $\bm B$ is rank equivalent to:
$$
 \bm{B'}= \left[ \begin{array}{cccc}-\Delta\bm p_{1,3} & \bm 0 & \bm 0& \bm 0 \\
  c_1 \Delta \bm p_{3,4} +c_2 \Delta \bm p_{3,5} & \bm 0 &\bm 0 & \bm 0\\
\bm 0& -\Delta \bm p_{2,4} & \bm 0 & \bm 0 \\
\Delta \bm p_{3,4} &-\Delta \bm p_{3,4} &\bm 0 &\bm  0 \\
\bm 0 & \bm 0 & -\Delta \bm p_{1,5}& \bm 0 \\
\Delta \bm p_{3,5} & \bm 0 & -\Delta \bm p_{3,5}& \bm 0\\
\bm 0& \Delta \bm p_{4,6} &\bm 0 & -\Delta \bm p_{4,6}\\
\bm 0& \bm 0 &\Delta \bm p_{5,6} & -\Delta \bm p_{5,6}
 \end{array}\right], 
$$
where $c_1=\frac{Det[\Delta \bm p_{4,6},\Delta  \bm p_{2,4}]\cdot Det[\Delta  \bm  p_{5,6},\Delta  \bm p_{2,6}]}{Det[\Delta  \bm p_{3,4},\Delta  \bm p_{2,4}]\cdot Det[\Delta  \bm p_{5,6},\Delta  \bm p_{4,6}]}$, $c_2=\frac{Det[\Delta  \bm p_{4,6},\Delta \bm  p_{2,6}]\cdot Det[\Delta  \bm p_{5,6},\Delta  \bm p_{1,5}]}{Det[\Delta \bm  p_{3,5},\Delta \bm  p_{1,5}]\cdot Det[\Delta  \bm p_{4,6},\Delta  \bm p_{5,6}]}$, and $Det[\bm i, \bm j]$ indicates the determinant of the $2\times 2$ matrix $\left[\footnotesize{\begin{array}{cc}\bm i \\ \bm j\end{array}}\right]$. The upper $2\times 2$ block of $\bm B'$ has determinant $c_1 Det[ \Delta \bm p_{3,4},\Delta \bm p_{1,3}]+c_2 Det[ \Delta \bm p_{3,5},\Delta \bm p_{1,3}]$. Because we assume under generic conditions that none of these vectors are parallel, each of the terms in this determinant are nonzero. Then, upon appropriate substitution and use of the geometric equality $Det([\bm i\;\;\bm j]^T)=|i| |j|\sin \theta$ (where $|\cdot|$ refers to the $2$-norm and $\theta$ is the angle formed between $\bm i$ and $\bm j$), we find this determinant is zero if and only if
\begin{align}
\frac{|\Delta \bm p_{1,5}|\sin \theta_1}{|\Delta \bm p_{2,6}|\sin \theta_2}=\frac{|\Delta \bm p_{1,3}|\sin \theta_3}{|\Delta \bm p_{2,4}|\sin \theta_4},
\label{eq:determinants}
\end{align}
where $\theta_1$ refers to the angle between $\Delta \bm p_{1,5}$ and $\Delta \bm p_{5,6}$, $\theta_2$ refers to the angle between $\Delta \bm p_{2,6}$ and $\Delta \bm p_{5,6}$, $\theta_3$ refers to the angle between $\Delta \bm p_{1,3}$ and $\Delta \bm p_{3,4}$, and $\theta_4$ refers to the angle between $\Delta \bm p_{2,4}$ and $\Delta \bm p_{3,4}$ (see Fig.~\ref{fig:quadrilateral}).

Under generic conditions in which none of the vectors in $\bm X_1\ast...\ast\bm X_5$ are parallel, the quantities on the left hand side of Eq.~\ref{eq:determinants} do not fully determine those on the right (see Fig.~\ref{fig:quadrilateral}). Therefore, Eq.~\ref{eq:determinants} is satisfied with probability $0$ and the upper left $2\times 2$ block in $\bm {B'}$ has full rank. Moreover, each $2\times 2$ block along the diagonal of $\bm{B'}$ has full rank. Therefore $\bm B$ itself has full rank, and the $15\times 18$ matrix $\bm X_1\ast... \ast \bm X_5$ has full row rank and right nullspace dimension $3$. If any one or more of the components $R_3$, $R_4$, and $R_5$ are singleton rods, then as in the end of the argument in Appendix B, we omit the corresponding point(s) $\bm p_{R_i}$ and the corresponding pair(s) of constraint rows, giving no net change in the determination of the degrees of freedom. 

\begin{figure}[t]
\begin{center}

\centering
\includegraphics[width=.6\linewidth]{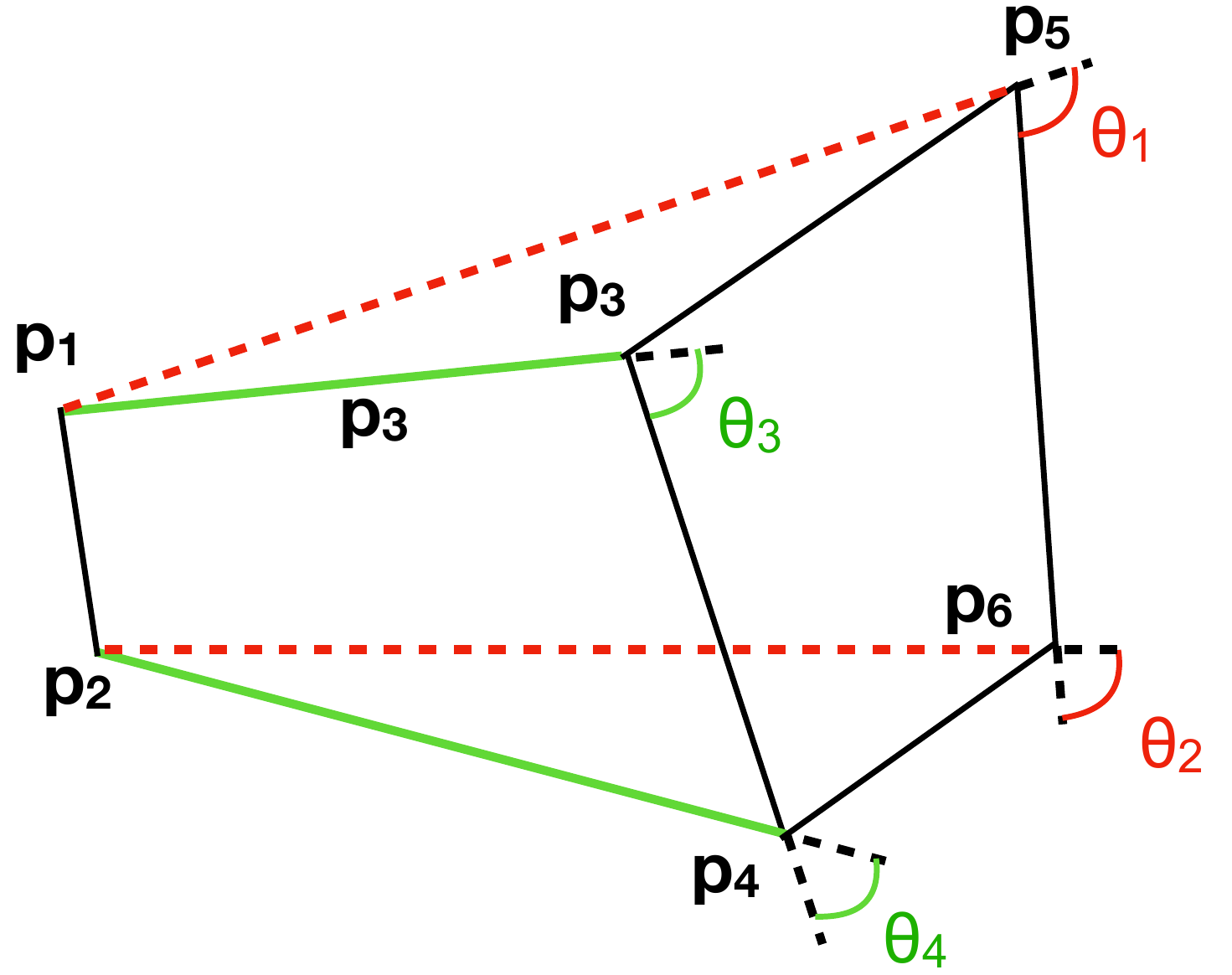}
\label{fig:2d51}
\caption{{\bf Visualization of quantities of interest in Eqs.~\ref{eq:determinants}.}
Given that $\bm p_1, \bm p_3, \bm p_5$ and $\bm p_2, \bm p_4, \bm p_6$ form noncollinear sets, the variables $|\Delta \bm p_{1,5}|$, $|\Delta \bm p_{2,6}|$, $\theta_1$, and $\theta_2$ (red) do not fully determine $|\Delta \bm p_{1,3}|$, $|\Delta \bm p_{2,4}|$, $\theta_3$, and $\theta_4$ (green).}
\label{fig:quadrilateral}
\end{center}
\end{figure}

If points $\bm p_1$, $\bm p_3$, and $\bm p_5$ lie along the same rod (as would be necessary if $R_1$ is a singleton rod), then the vectors $\Delta \bm p_{1,5}$ and $\Delta \bm p_{3,5}$ are parallel and the matrix $\bm B$ defined above does not have full row rank. Above, we excluded this possibility; now we turn to the case in which the sets $\{\bm p_1, \bm p_3, \bm p_5\}$ and $\{ \bm p_2, \bm p_4, \bm p_6\}$ are both collinear sets (i.e., each set of three intersection points is along a single rod). This collinearity condition makes our task more challenging, as each of the three \emph{collinear} points in $\{\bm p_1, \bm p_3, \bm p_5\}$ (and again in $\{ \bm p_2, \bm p_4, \bm p_6\}$) lie in $R_1$ ($R_2$), yet we must choose three \emph{noncollinear} points for a coordinate labeling. First, we choose without loss of generality to label $\bm p_3$ to lie between $\bm p_1$ and $\bm p_5$, and $\bm p_4$ to lie between $\bm p_2$ and $\bm p_6$. Then, we choose as coordinate labelings $\{\bm p_1, \bm p_5,\bm p_{R_1}\}$ for $R_1$, and $\{\bm p_2, \bm p_6,\bm p_{R_2}\}$ for $R_2$, where $\bm p_{R_1}$ is chosen to be noncollinear with $\bm p_1$ and $\bm p_5$ (and, similarly, $\bm p_{R_2}$ with $\bm p_2$ and $\bm p_6$). We choose coordinate labelings $R_3, R_4$, and $R_5$ as before.

This choice of labelings give $15$ constraints as above, but we have now included $11$ points in these constraints, so even if these constraints are linearly independent, there appear to be $2(11)-15=7$ degrees of freedom remaining. However, as alluded to in Sec.~\ref{sec:previous_discussion}, the  constraints of the form in Eq.~\ref{eq:rigidity_row} do not account for the fact that all points along a single rod are rigid with respect to another. Given that $\bm p_3$ lies between $\bm p_1$ and $\bm p_5$, we account for this constraint explicitly using the geometric equality
\begin{equation} 
\bm p_3=s \bm p_5 +(1-s)\bm p_1,
\end{equation} 
where $s=\frac{|\Delta \bm p_{1,3}|}{|\Delta \bm p_{1,5}|}\in(0,1)$. We can think of this equation parametrically, with $s$ giving the fractional distance along the line segment from $\bm p_1$ to $\bm p_5$ at which $\bm p_3$ is located. Because this fractional distance is fixed (i.e., $\bm p_3$ cannot shift along the rod), we also have the condition that $\frac{ds}{dt}=0$, and therefore
\begin{equation} 
\bm u_3=s \bm u_5 +(1-s)\bm u_1\,.
\end{equation}
We can similarly derive that $\bm u_4=s' \bm u_6+(1-s')\bm u_2$, where $s'=\frac{|\Delta \bm p_{2,4}|}{|\Delta \bm p_{2,6}|}\in(0,1)$. Each equation introduces two \emph{augmented} constraints into the augmented $19\times 22$ composite rigidity matrix. Using steps similar to those detailed in the proofs above, we can then show that this latter matrix has full row rank unless $\Delta \bm p_{1,2}$, $\Delta \bm p_{3,4}$, and $\Delta \bm p_{5,6}$ are mutually parallel (which we excluded by hypothesis). The case in which one but not both of the sets $\{\bm p_1,\bm p_3,\bm p_5\}$ and $\{\bm p_2,\bm p_4,\bm p_6\}$ are collinear follows similarly.

\section{Intersections of Motifs 2D2, 2D3, and 2D5}\label{app:order}
In general, iterative graph compression of two intersecting motifs might achieve different final states if different orderings of motif compression are used. However, in testing the motifs used here on small systems (described below), we do not identify any cases where the output of \emph{2D-RGC-5} or \emph{2D-RGC-3} is affected by the order of compression of 2-, 3-, and 5-component motifs (as well as 3-clique communities). 

First, we consider simple cases in which the different motifs share at least one node in common. For example, suppose a 2-component and a 3-component motif intersect. There are two ways in which this may occur: the former may be fully contained in the latter, or the motifs may simply share a node (see the top two graphs in Fig.~\ref{fig:order}). In the former case, application of Motif 2D2 yields another 2D2 motif (which is then compressed), while application of Motif 2D3 compresses the graph in a single step. In the latter case, application of Motif 2D3 leads to Motif 2D2, and vice versa. In Fig.~\ref{fig:order}, we enumerate the possible (nonisomorphic) ways in which any of the 2-, 3-, and 5-component motifs may intersect pairwise. 
It is easy to check that any order of motif compression will yield a single rigid component in these cases. 

We have additionally employed the graphlet-based exhaustive search method used in Sec.~\ref{sec:results} to verify that each ordering of graph compression gives the same output for all rod contact networks generated with up to $r=8$ rods. Generally, we thus expect that graph compression ordering is very likely inconsequential in two dimensions, but we believe it may have some importance for 3D rigid graph compression.
\begin{figure}[t]
\begin{center}
\includegraphics[width=.5\linewidth]{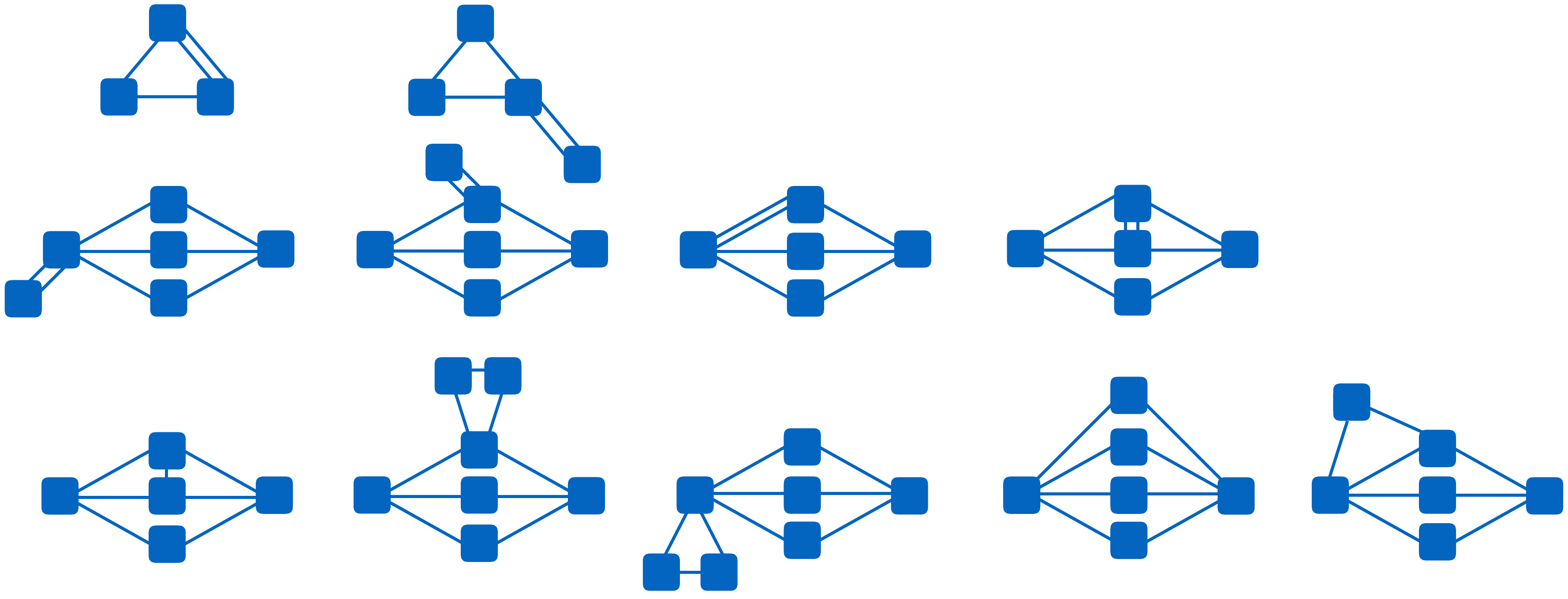}
\caption{{\bf Nonisomorphic intersections of 2-, 3-, and 5-component motifs.}
Top Row: Simplest nonisomorphic cases intersecting (that is, containing both) the 2- and 3-body primitive rigid motifs.
Middle Row: Simplest intersections of 2- and 5-component motifs. 
Bottom Row: Simplest intersections of 3- and 5-component motifs. 
Each of these networks compress to a single rigid component regardless of the order in which the 2D primitive rigid motifs are compressed.}
\label{fig:order}
\end{center}
\end{figure}

\bibliography{references}
\bibliographystyle{siam}

\end{document}